\documentclass[]{aa} 

\usepackage{graphicx}
\usepackage{txfonts}
\usepackage{natbib}
\usepackage{longtable}

\begin{document}

\title{A Corona Australis cloud filament seen in NIR scattered light}
\subtitle{III. 
Modelling and comparison with Herschel sub-millimetre data\thanks{{\it Herschel} is an ESA space
observatory with science instruments provided by European-led
Principal Investigator consortia and with important participation from
NASA.}}

\author{M.     Juvela\inst{1},
        V.-M.  Pelkonen\inst{1,2},
        G. J.  White\inst{3,4},
        V.     K\"onyves\inst{5},
        J.     Kirk\inst{6},
        P.     Andr\'e\inst{5}
        }

\institute{
Department of Physics, P.O.Box 64, FI-00014, University of Helsinki,
Finland, {\em mika.juvela@helsinki.fi}
\and
Finnish Centre for Astronomy with ESO (FINCA), University of Turku,
V\"ais\"al\"antie 20, FI-21500 Piikki\"o, Finland
\and
Department of Physics and Astronomy, The
Open University, Walton Hall, Milton Keynes, MK7 6AA, UK
\and
RAL
Space, STFC
Rutherford Appleton Laboratory, Chilton, Didcot, Oxfordshire,
OX11 0QX, UK
\and
Laboratoire AIM, CEA/DSM--CNRS--Universit\'e Paris Diderot,
IRFU/Service d'Astrophysique, CEA Saclay, 91191 Gif-sur-Yvette, France
\and
School of Physics and Astronomy, Cardiff University, Queen's
Buildings, Cardiff CF24 3AA
}

\authorrunning{M. Juvela et al.}

\date{Received September 15, 1996; accepted March 16, 1997}

\abstract
{
The dust is an important tracer of dense interstellar clouds but its
properties are expected to undergo changes affecting the scattering
and emitting properties of the grains. With recent Herschel
observations, the northern filament of the Corona Australis cloud has
now been mapped in a number of bands from 1.2\,$\mu$m to 870\,$\mu$m.
The data set provides a good starting point for the study of the cloud
over several orders of magnitude in density.
}
{
We wish to examine the differences of the column density distributions
derived from dust extinction, scattering, and emission, and to
determine to what extent the observations are consistent with the
standard dust models.
}
{
From Herschel data, we calculate the column density distribution that
is compared to the corresponding data derived in the near-infrared
regime from the reddening of the background stars, and from the surface
brightness attributed to light scattering. We construct
three-dimensional radiative transfer models to describe the emission
and the scattering. 
}
{
The scattered light traces low column densities of $A_{\rm
V}\sim$1$^{\rm m}$ better than the dust emission, remaining useful
to $A_{\rm V}\sim 10-15^{\rm m}$.
Based on the models, the extinction and the level of dust emission are
surprisingly consistent with a sub-millimetre dust emissivity typical of diffuse
medium. However, the intensity of the scattered light is very low at
the centre of the densest clump and this cannot be explained without a
very low grain albedo. Both the scattered light and dust emission
indicate an anisotropic radiation field. The modelling of the dust
emission suggests that the radiation field intensity is at least three
times the value of the normal interstellar radiation field.
}
{
The inter-comparison between the extinction, light scattering, and
dust emission provides very stringent constraints on the cloud
structure, the illuminating radiation field, and the grain properties.
}

\maketitle
%

\section{Introduction}

The interstellar clouds are hierarchical structures where 
gravitationally bound prestellar cores are found on the smallest
scales. The study of the dense clouds is largely motivated by this
connection to star formation. 
The filamentary nature of interstellar clouds and the possible
connection between filaments and star formation has been known for a
long time \citep[e.g.][]{Barnard1919, Fessenkov1952, Elmegreen1979,
Schneider1979, Bally1987}.
The recent ground-based studies and Herschel surveys have shown cloud
filaments in exquisite detail \citep{MAMD2010, Andre2010,
Arzoumanian2011, Hill2011, Juvela2010} and have demonstrated that
pre-stellar cores and protostars are preferentially located along
these structures \citep{Andre2010, Menshchikov2010, PaperIII}.
Filaments are a natural outcome of interstellar turbulence
\citep[e.g.][]{PadoanNordlund2011}, with further contributions from
immediate triggering by supernova explosions and the radiation and
stellar winds from massive stars.
However, some filaments may also be formed directly by gravitational
processes, as have been modelled in numerous cosmological simulations,
and in other theoretical studies \cite[e.g.][]{Burkert2004}.
The filaments should fragment as dictated by the local Jeans
condition, a process addressed by many theoretical studies
\citep{Inutsuka1997, Myers2009}.  In addition to the pure compressive
instability, also the so-called sausage instability may sometimes play
an additional role \citep[see][]{Fischera2012, McLeman2012}. Although
the morphology of {\em all} the star-forming clouds is not
predominantly filamentary \citep{PaperIII}, a global picture of star
formation is emerging, where the turbulence creates filaments, the
filaments become gravitationally unstable and subsequently fragment
forming the cores that may still fed by material flowing in along the
filaments. Low mass stars can be found in individual filaments while
high mass stars are preferentially born at the intersections of
several filaments. This scenario is supported by the observations made
within the Herschel Gould Belt survey \citep[e.g.][]{Andre2010,
Konyves2010, Menshchikov2010} and other Herschel programs
\citep{Molinari2010, Schneider2010, NguyenLuong2011, Hill2011,
Schneider2012, PaperIII} as well as by numerical simulations
\citep{PadoanNordlund2011, VazquezSemadeni2011, Klessen2011,
Bonnell2011}. 

To determine the initial conditions for the star formation to occur,
we need to measure the physical properties of the clouds, the
filaments, and the cores. The mass distributions of the clouds can be
measured with a number of methods. This is fortunate because each
method suffers from different sources of uncertainty.

The dense clouds have been traditionally mapped in molecular lines.
Line observations are invaluable because of the information they give
on the physical state, kinematics, and chemistry of the clouds
\citep{Bergin2007}. On the other hand, they do not always provide reliable
or consistent estimates of the mass distribution. The line intensity depends on the
local physical conditions, mainly density and temperature, and the
chemical abundances. If the lines are not optically thin, the
radiative transfer effects introduce additional uncertainty. 

For the above reasons -- and because of the development in detector
technology and the appearance of new ground-based, balloon-borne, and
space-borne facilities -- the thermal dust emission has become
important as a tracer of the densest clouds. Observations of the dust
at submm wavelengths are of particular interest because they are
sensitive to the emission of the cold dust that may have been missed
in earlier far-infrared studies. With the knowledge of dust
temperature and dust opacity, the column density can be estimated.
However, the colour temperature obtained from observations is known to
be a biased estimator of the average dust temperature
\citep{Shetty2009a, Juvela2012_TB} and additionally the value of the
dust opacity is uncertain. Furthermore, there are clear indications
that the dust opacity is not constant, the variations being a likely 
consequence of grain coagulation and aggregation processes
\citep{Cambresy2001, delBurgo2003, Kramer2003, Lehtinen2007}.
Similarly, the spectral index appears to vary from region to region
and this increases the uncertainty of the colour temperature and
column density estimates. The observations point to a negative
correlation between the colour temperature $T_{\rm d}$ and the
observed spectral index $\beta_{\rm Obs}$ \citep{Dupac2003,
Desert2008, Anderson2010, Paradis2010, Veneziani2010, PlanckI,
Arab2012}. Such variations can affect the accuracy of the column
density estimates derived from the dust emission. However, the
evaluation of errors is complicated by the observational noise (also
through its effect on the $T_{\rm d}-\beta_{\rm Obs}$ relation, see
\citet{Shetty2009b, Juvela2012_CHI2} and the unknown line-of-sight
temperature variations \citep{Shetty2009a, Juvela2012_CHI2}. One of
the main advantages of submm observations of dust emission is the
large range of column densities probed. With the Herschel satellite
SPIRE instrument \citep{Griffin2010}
one can probe regions from $A_{\rm V}\sim 1^{\rm m}$ to $A_{\rm V}\sim
100^{\rm m}$ with spatial resolution of $\sim 30\arcsec$ . However,
the accuracy will depend on the line-of-sight temperature variations
and other opacity effects \citep{Malinen2011}.

Complementary information can be obtained from near-infrared (NIR)
observations, from reddening of the light of background stars, and
from the measurement of the scattered light. The observed
near-infrared colour excesses can be converted to column density or to
provide estimates of the visual extinction $A_{\rm V}$ using
techniques such as the NICER algorithm \citep{Lombardi2001}. The
calculations are based on the assumption that the average intrinsic
colour of the stars is constant and, therefore, reliable estimates are
obtained only as an average over a number of stars. Therefore, the
extinction must be estimated for larger pixels, as an average over
$\sim$10 or more background stars. This reduces the resolution of the
extinction maps and can cause bias in the presence of strong $A_{\rm
V}$ gradients. With dedicated observations it is possible to achieve a
spatial resolution of some tens of arc seconds, which is comparable to
the resolution of the submm surveys. The range of $A_{\rm V}$
values probed extends from $A_{\rm V}\sim1^{\rm m}$ to $\sim40^{\rm
m}$, depending on the stellar density and the depth of the
observations.

If the NIR observations are made using an ON-OFF mode the diffuse
signal is preserved and it is possible to measure the light scattered
by the same dust particles that are responsible for the NIR
extinction. If the cloud is optically thin at the observed wavelengths
and if the cloud does not include local radiation sources, the signal
will be proportional to the column density \citep{Juvela2006_SCA}. The
main advantage of the surface brightness measurement is the high,
potentially $\sim 1\arcsec$ spatial resolution. Even in normal NIR
observations at mid-Galactic latitudes, the scattered light will
provide better resolution than similar observations of the background
stars. The method remains useful in the range $A_{\rm V}=1$--$15^{\rm
m}$. At higher column densities the signal saturates and the surface
brightness will depend on the structure of the cloud. With
multiwavelength measurements (e.g., J, H, and K bands) it becomes
possible to use the intensity ratios to measure and to partially
correct for the saturation of the signal \citep[][hereafter Paper I and Paper
II, respectively]{Juvela2006_SCA, Juvela2008_CrA}. The observed
surface brightness depends on the grain properties and upon the
intensity of the radiation field at the location of the cloud.
Therefore, combining surface brightness data with other column density
tracers one can also provide constraints to these parameters.

In this paper we will examine a filament in the northern end of the
Corona Australis molecular cloud. The filament contains a dense clump
with a central $A_{\rm V}$ previously estimated to be at least
$\sim30^{\rm m}$. The source has been studied with the help of NIR
observations using background stars and the scattered light
\citep{Juvela2006_SCA, Juvela2008_CrA, Juvela2009_CrA}. The
region has now been mapped with the Herschel satellite and these data
provide the opportunity to measure the structure of the central part
of the filament where the column density was too high to enable reliable
estimates either with the NICER or with the surface brightness (scattered
light) methods. This provides an excellent opportunity to compare the
performance of the three column density tracers and to use the
comparison to look for indications for changes in the dust properties.

The structure of the paper is the following: In
Sect.~\ref{sect:observations} we describe the observations and the
main properties of the data. In Sect.~\ref{sect:results} we present
the direct results derived from each data set and make the first
comparison between the column density estimates. We continue by
constructing radiative transfer models for the NIR data and the
sub-millimetre data separately. These calculations are described and
compared in Sect.~\ref{sect:modelling}. Our final conclusions are
given in Sect.~\ref{sect:conclusions}.

\begin{figure*}
\centering
\includegraphics[width=16cm]{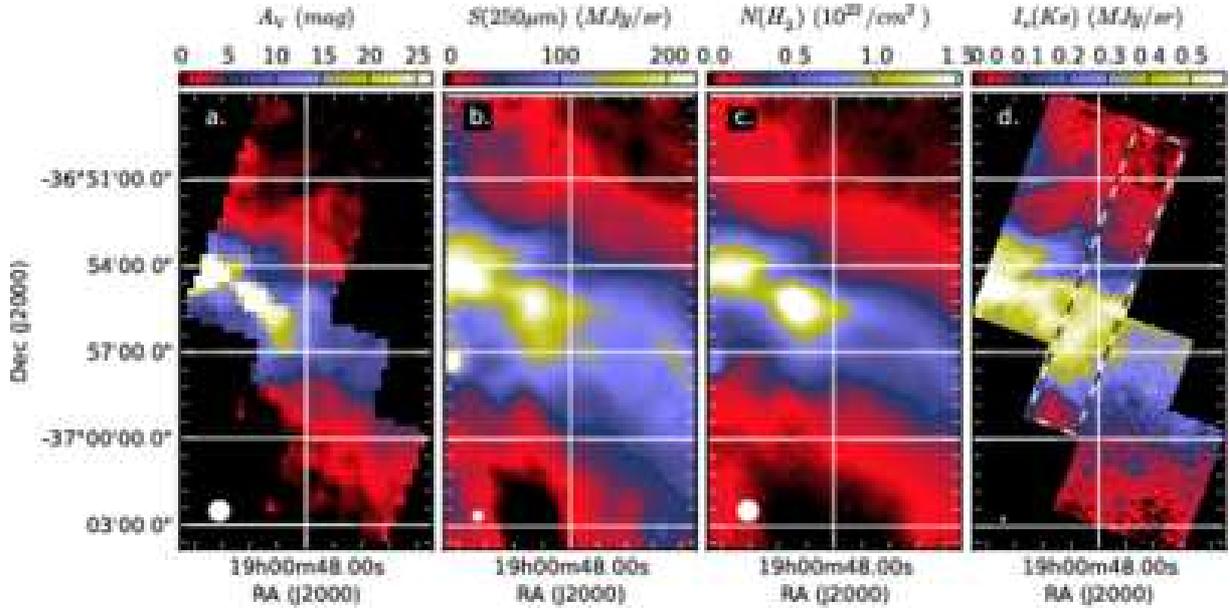}
\caption{
A cross section of the northern Corona Australis filament according to
the different tracers. Maps are shown for the visual extinction
derived from the reddening of the background stars (frame $a$,
40$\arcsec$ resolution), the 250\,$\mu$m surface brightness (frame
$b$, 18$\arcsec$ resolution), the column density derived from the
Herschel data (frame $c$, 40$\arcsec$ resolution), and the intensity
of the scattered light in the $K_s$ band (frame $d$, 4$\arcsec$
resolution).  For the plot the data have been adjusted so that the
median signal is zero within the reference area located in the
northern end of the maps (see text). The dashed rectangle in
frame $d$ denotes the area that will be used to measure the filament
profile.
}
\label{fig:pretty_plot}%
\end{figure*}

\section{The observations} \label{sect:observations}

\subsection{Near-infrared data} \label{sect:obs-NIR}

The NIR observations were made with the SOFI instrument on the NTT
telescope, Chile. All observations were carried out in ON-OFF mode. 
The northern part, an area of $4\arcmin \times 8\arcmin$ was observed
in 2006. The integration times on the ON-fields were 1.0 hour in the J
band, 1.5 hours in the H band, and 4.5 hours in the Ks band. The map
was extended in June 2007 with two $4\arcmin \times 4\arcmin$ fields
that cover the southern side of the filament. The integration times
were 1.5 hours for the Ks band and 0.5 hours in the J and H bands. On
the basis of the 2MASS data \citep{Skrutskie2006}, the OFF field was
estimated to have an extinction below $A_{\rm V}=0.5^{\rm m}$.  For
the details of the observations, see \citet{Juvela2008_CrA} and
\citet{Juvela2009_CrA}.

The observations were calibrated using the 2MASS stellar photometry
and that calibration was carried over to the surface brightness. For
the final surface brightness map the southern ON fields were mosaiced
together with the northern part, this requiring a small additive
adjustment. The observations do not provide an absolute measure of the
surface brightness. Therefore, we present the results relative to the
signal within a $1\arcmin$ diameter aperture centred at coordinates
$\alpha_{\rm 2000}$=19$^{\rm h}$0$^{\rm m}$43.5$^{\rm s}$,
$\delta_{\rm 2000}$=-36$\degr$49$\arcmin$36$\arcsec$ (see
Fig.~\ref{fig:pretty_plot}d). Based on the 2MASS data, the visual
extinction in this reference area is $\sim 1.8^{\rm m}$
\citep[see][]{Juvela2009_CrA}. 

In \citet{Juvela2008_CrA} the NICER method \citep{Lombardi2001} was
used to convert the measured stellar colour excesses to estimates of
extinction. This extinction map is reproduced in
Fig.~\ref{fig:pretty_plot}a with $A_{\rm V}$ values relative to the
above described reference area. The frames b and c show for the same
area the 250\,$\mu$m surface brightness and the column density derived
from the Herschel data.

\subsection{Herschel observations} \label{sect:obs-submm}

The Corona Australis cloud has been mapped as part of the Herschel
Gould Belt Survey \citep{Andre2010}. The entire cloud is covered by
parallel mode maps that also cover the northern filament. The
observation were made with the PACS instrument\footnote{PACS has been developed
by a consortium of institutes led by MPE (Germany) and including UVIE
(Austria); KU Leuven, CSL, IMEC (Belgium); CEA, LAM (France); MPIA
(Germany); INAF-IFSI/OAA/OAP/OAT, LENS, SISSA (Italy); IAC (Spain).
This development has been supported by the funding agencies BMVIT
(Austria), ESA-PRODEX (Belgium), CEA/CNES (France), DLR (Germany),
ASI/INAF (Italy), and CICYT/MCYT (Spain).} \citep{Poglitsch2010}
at wavelengths 70\,$\mu$m and 160\,$\mu$m and with the 
SPIRE\footnote{SPIRE
has been developed by a consortium of institutes led by Cardiff
University (UK) and including Univ. Lethbridge (Canada); NAOC (China);
CEA, LAM (France); IFSI, Univ. Padua (Italy); IAC (Spain); Stockholm
Observatory (Sweden); Imperial College London, RAL, UCL-MSSL, UKATC,
Univ. Sussex (UK); and Caltech, JPL, NHSC, Univ. Colorado (USA). This
development has been supported by national funding agencies: CSA
(Canada); NAOC (China); CEA, CNES, CNRS (France); ASI (Italy); MCINN
(Spain); SNSB (Sweden); STFC (UK); and NASA (USA)}
instrument \citep{Griffin2010} at wavelengths 250\,$\mu$m, 350\,$\mu$m,
and 500\,$\mu$m. The observation consist of two orthogonal scans with
observation identifiers 1342206677 and 1342206678. We use data
provided by the Herschel Gould Belt Consortium.
We use the SPIRE maps produced with the naive mapping routine in HIPE 
software \citep{Ott2010}. The 160\,$\mu$m PACS map is made using
Scanamorphos v. 16 (Roussel 2011, submitted), starting from data
processed to level 1 in HIPE. The 70\,$\mu$m data were not used, since
at that wavelength the emission is not produced by the large dust
grains that are responsible for the longer wavelength emission. For
the Herschel data the zero point of the intensity scale was estimated
by comparison with data from the Planck satellite survey
\cite[see][]{Bernard2010}. These zero points are used in the
derivation of the column density estimates. However, in the plots
(e.g., Fig.\ref{fig:pretty_plot}b) and in comparison with the other
tracers, the zero level is set using the reference area described in
Sect.~\ref{sect:obs-NIR}.

\subsection{Laboca 870\,$\mu$m data} \label{sect:obs870}

The Corona Australis filament was mapped in 2008 with the APEX/LABOCA
instrument. This instrument operates at a wavelength of 870\,$\mu$m,
and has the beam size $\sim 19\arcsec$. Observations consist of $\sim
30 \arcmin$ long scans that were made perpendicular to the filament,
and at up to 30 degree angles relative to the normal of the filament.
The observations were reduced using the Boa program, version 1.11 and
the calibration checked with planet observations and with the help of
the source S CrA that is located immediate south-east of the area
covered by the NIR data. The final noise level in the central part of
the map is better than 20\,mJy/beam. For details of the Laboca
observations, see \citet{Juvela2009_CrA}.

\subsection{Mid-infrared data} \label{sect:spitzer}

\begin{figure*}
\centering
\includegraphics[width=15.3cm]{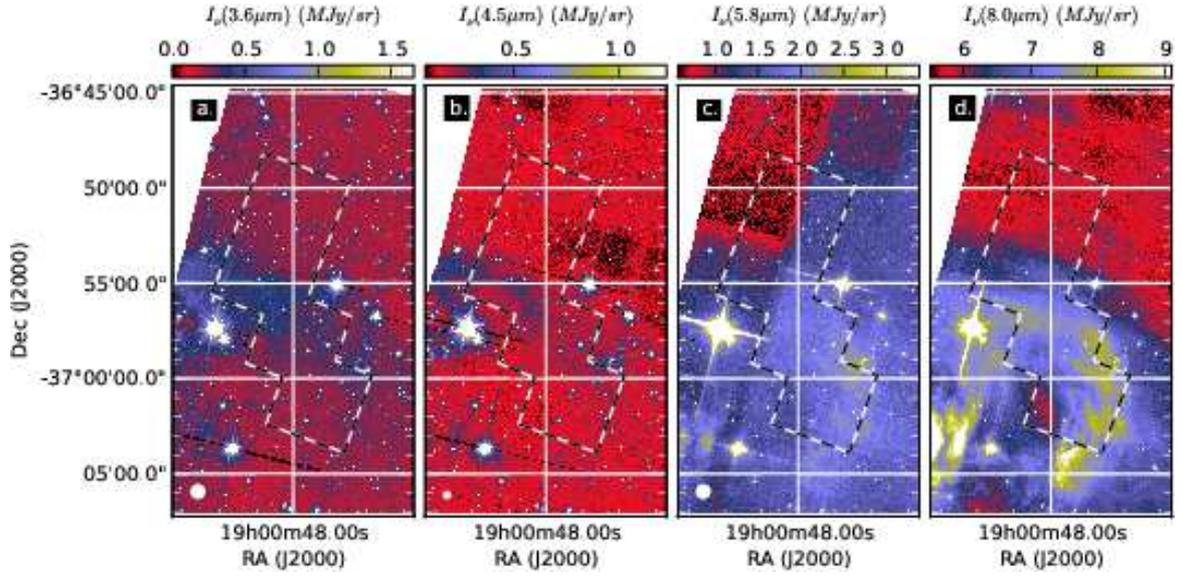}
\caption{Spitzer IRAC maps at 3.6, 4.5, 5.8, and 8.0\,$\mu$m. The
dashed line shows the outline of the regions covered by NIR observations.
The brightest star east of the area covered by NIR observations
is S CrA. The other bright source in the southern end of the images
(at $\delta_{\rm 2000}=-37\degr3\arcmin42\arcsec$) corresponds to the blended image
of the stars HD\,176269 and HD\,176270.} 
\label{fig:IRAC}
\end{figure*}

The area was mapped in Spitzer guaranteed time programs with both
the IRAC instrument (3.6, 4.5, 5.8, and 8.0\,$\mu$m) and the MIPS
instrument (24, 70, and 160\,$\mu$m). 
The MIPS data were already compared to the LABOCA and NIR data in
\cite{Juvela2009_CrA}. The images of the four IRAC bands are shown in
Fig.\ref{fig:IRAC}. The 5.8 and 8.0\,$\mu$m surface brightness is
affected by the nearby stars and does not trace the column density.
The 3.6 and 4.5\,$\mu$m data will be used in
Sect.~\ref{sect:coreshine} check for signs of enhanced mid-infrared
scattering.

We will use also data from the WISE\footnote{The Wide-field Infrared
Survey Explorer is a joint project of the University of California,
Los Angeles, and the Jet Propulsion Laboratory/California Institute of
Technology, funded by the National Aeronautics and Space
Administration.} survey that have recently become public. The WISE
data includes the wavelengths of 3.4, 4.6, 12, and 22\,$\mu$m
\citep{Wright2010}.

\section{The results}  \label{sect:results}

\subsection{The column densities}
\label{sect:column_densities}

\subsubsection{Column density derived from the Herschel data}
\label{sect:results-submm}

The Herschel measurements of submm dust emission were
converted to column density using the usual procedure that assumes
a constant dust temperature along the line of sight and a constant
dust opacity everywhere within the mapped area. For each map pixel, the
observed intensities $I_{\nu}$ at 160\,$\mu$m, 250\,$\mu$m,
350\,$\mu$m, and 500\,$\mu$m were fitted with a modified black body
curve $B_{\nu}(T_{\rm C})\nu^{\beta}$ to determine the colour
temperature $T_{\rm C}$.  With knowledge of the dust opacity
$\kappa$ (and the spectral index $\beta$) the fit determines
the column density. The column density of the molecular hydrogen can
be written formally as:
\begin{equation}
    N({\rm H}_2) = \frac{ I_{\nu} }{ B_{\nu}(T) \kappa \mu m_{\rm H}},
    \label{eq:colden}
\end{equation}
where $T$ is the assumed temperature, here equal to the colour
temperature $T_{\rm C}$, $\mu$ is the average particle mass per
hydrogen molecule, and the dust opacity $\kappa$ is given relative to
the gas mass.

In the analysis, the surface brightness maps are convolved to a common
resolution of $40\arcsec$ resolution before the determination of
$T_{\rm C}$ and $N$(H$_2$). The spectra are modelled assuming a
constant value of the spectral index, $\beta=2.0$ and a dust opacity
$\kappa=$0.1\,cm$^2$/g\,($\nu$/1000\,GHz)$^{\beta}$ that is applicable
to high density environments \citep{Hildebrand1983,
Beckwith1990}. The same opacity law has been adopted in several papers
on Herschel results \citep{Andre2010, Konyves2010, Arzoumanian2011,
PlanckII, PaperIII}. However, for any particular sources, the values
of these parameters are not known to high precision and both the
$\beta$ and the $\kappa$ values introduce an uncertainty in the column
density that may be several tens of per cent.

\begin{figure}
\centering
\includegraphics[width=6.5cm]{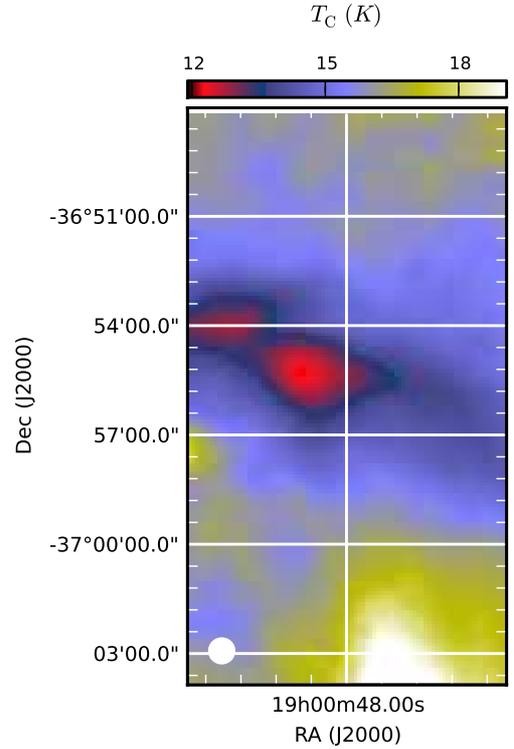}
\caption{The map of the dust colour temperature $T_{\rm C}$.}
\label{fig:TC}
\end{figure}

The calculated column density map is shown in
Fig.~\ref{fig:pretty_plot}c and the colour temperature map is shown
separately in Fig.~\ref{fig:TC}. The minimum temperature at the
location of the densest clump is $T_{\rm C}\sim 12.5$\,K. This value
is obtained using the total intensity (i.e., the original zero point
of the Herschel maps) without subtracting the signal in the local
reference area (see Sect.~\ref{sect:obs-submm}).

\begin{figure*}
\centering
\includegraphics[width=16cm]{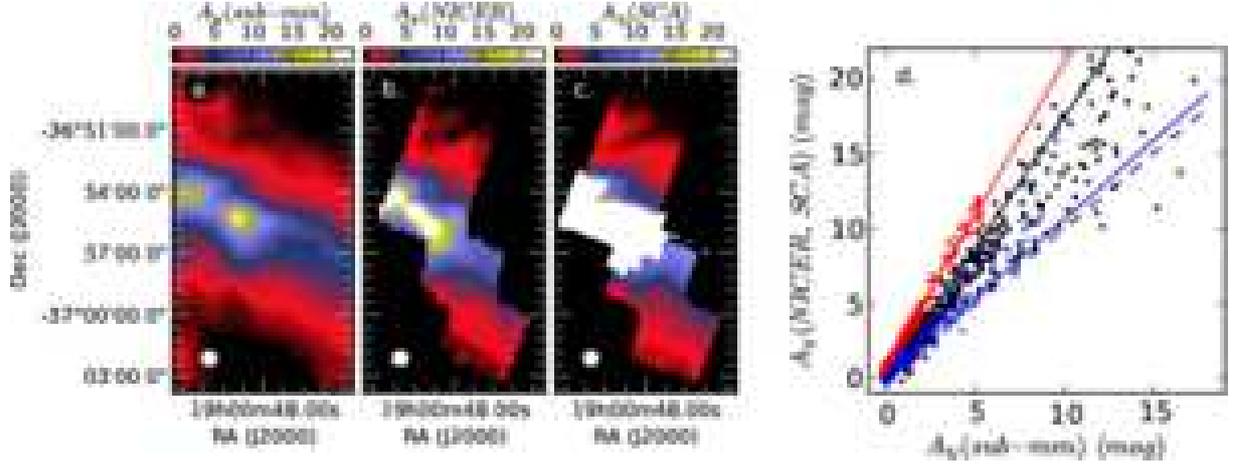}
\caption{
A comparison of the $A_{\rm V}$ estimates derived from the
sub-millimetre emission (frame $a$, the column density converted to
$A_{\rm V}$ with the assumptions listed in the text), the reddening of
the background stars (frame $b$), and the dust scattering (frame $c$).
All maps have been convolved to a common resolution of 40$\arcsec$.
The frame $d$ shows the NICER $A_{\rm V}$ estimates (black
points) and the estimates $A_{\rm V}^{\rm SCA}$ derived from the NIR
surface brightness as the function of the values from sub-millimetre
data. The blue and the red points are, respectively, the $A_{\rm
V}^{\rm SCA}$ values north and south of declination $-$36$\degr$55$\arcmin$. The
solid lines are least squares fits to the respective data points. The
dashed line indicates the one-to-one relation.
}
\label{fig:colden_comparison}
\end{figure*}

\subsubsection{The $A_{\rm V}$ maps from NIR data}

The extinction derived from the colour excesses of the background
stars was calculated using the NICER method \citep{Lombardi2001}. The
resulting maps of visual extinction, $A_{\rm V}^{\rm NICER}$ have a
resolution of 30$\arcsec$. A similar map with a 20$\arcsec$ resolution
has already been presented by \cite{Juvela2008_CrA}. Here we settle for a
more reliable but a slightly lower initial resolution of 30$\arcsec$
because the data will later be compared to lower resolution column
density maps derived from the Herschel data. The extinction map covers
the whole area included in the NIR observations (see
Fig.\ref{fig:pretty_plot}a). The maximum extinction values are
$\sim$33$^{\rm m}$ but because no background stars are visible at the
centre of the dense clumps within the filament, this is only a lower
limit. Furthermore, because of the strong $A_{\rm V}$ gradients the
values near the filament centre may be biased downwards
\citep[see][]{Juvela2008_CrA}.

In \cite{Juvela2008_CrA} and \cite{Juvela2009_CrA} the surface
brightness, assumed to consist only of scattered light, was converted
to column density assuming analytical formulas $I_{\nu} \sim
a_{\nu}(1-exp(-b_{\nu} A_{\rm V}))$ for the relation between the surface
brightness and the extinction. This resulted in an extinction map
$A_{\rm V}^{\rm SCA}$ that was consistent with the $A_{\rm V}^{\rm
NICER}$ values up to $\sim 10^{\rm m}$ but gave higher values in the
range of 10--15$^{\rm m}$. It was not clear which of the two
extinction estimates was more robust although simulations indicated
that most of the difference could be explained by bias in $A_{\rm
V}^{\rm NICER}$. When the extinction approaches 20$^{\rm m}$ the
surface brightness is strongly saturated and the simple analysis
method adopted in \cite{Juvela2008_CrA, Juvela2009_CrA} leads to
diverging values. The saturation of the surface brightness and the dip
at the centre of the dense clump are evident even in the $K_s$ band
(see Fig.\ref{fig:pretty_plot}d) where, compared to the $V$ band the
optical depth is lower by close to a factor of ten. This alone 
confirms that the central $A_{\rm V}$ must be higher than 20$^{\rm
m}$. As in \cite{Juvela2008_CrA}, the spatial resolution of this
initial $A_{\rm V}^{\rm SCA}$ map is 10$\arcsec$.

\subsubsection{The comparison of the column density maps}
\label{sect:compare_colden}

In Fig.~\ref{fig:colden_comparison} we compare the column density
estimates from the submm emission, the reddening of the background
stars (the NICER method by \cite{Lombardi2001}), and from the NIR
surface brightness data \citep{Juvela2009_CrA}. 

For easier comparison, the Herschel estimates of $N$(H$_2$) were also
converted to units of visual extinction, $A_{\rm V}^{\rm sub-mm}$. The
conversion was done using the canonical relation
$N$(H$_2$)=$0.94\times 10^{21} A_{\rm V}$ \citep{Bohlin1978}. That
relation was determined for more diffuse lines of sight and may not be
representative of the present field, but does at least provide a
convenient point of reference. The map of $A_{\rm V}^{\rm sub-mm}$ (a
scaled version of Fig.~\ref{fig:pretty_plot}c) is shown in
Fig.~\ref{fig:colden_comparison}a.
The map of $A_{\rm V}^{\rm NICER}$ was convolved to the resolution of
40$\arcsec$. We have masked the areas near the borders of the observed
area where the result of the convolution is not well defined (where
more than 10\% of the convolving beam falls outside observed region).
The $A_{\rm V}^{\rm NICER}$ map is shown in
Fig.~\ref{fig:colden_comparison}b.
In the case of the NIR surface brightness, $A_{\rm V}^{\rm SCA}$
values remain undefined in the central part of the filament where the
simple analytical solution diverges. In the area where the $A_{\rm
V}^{\rm SCA}$ estimates exceed 25$^{\rm m}$, before making the
convolution, the data were replaced with values read from the NICER
map. After the convolution, those areas are masked from the subsequent
analysis (Fig.~\ref{fig:colden_comparison}c).
In the plots Fig.~\ref{fig:colden_comparison}a--c the map zero points
were set using the common reference region (see \ref{sect:obs-NIR}).

The three $A_{\rm V}$ maps are morphologically very similar at low and
intermediate extinctions. In the lowest $A_{\rm V}$ regions the
absolute value of the visual extinction is of order of 1--3$^{\rm
mag}$. At those diffuse lines-of-sight the correspondence is best
between the two NIR-derived maps because one is close to the
sensitivity of the sub-millimetre data (or because the sub-millimetre
maps may be affected by minor baseline uncertainties).

Figure~\ref{fig:colden_comparison}d confirms the good correspondence
between the tracers.  Below $A_{\rm V}^{\rm sub-mm}=10^{\rm m}$ the
correlation coefficients coefficient is 0.96 for the $A_{\rm
V}^{NICER}$--$A_{\rm V}^{\rm sub-mm}$ relation with a least squares
line $A_{\rm V}^{NICER}$=-0.11+1.76$\times A_{\rm V}^{\rm sub-mm}$.
For the relation $A_{\rm V}^{SCA}$--$A_{\rm V}^{\rm sub-mm}$ the
correlation coefficient is $r=$0.86 but the presence of two different
relations is evident. In the northern half of the map (blue points in
Fig.~\ref{fig:colden_comparison}d) the least squares fit gives a
relation $A_{\rm V}^{SCA}$=0.15+1.05$\times A_{\rm V}^{\rm sub-mm}$
with a correlation coefficient of $r=$0.95. The slopes are consistent
with the results presented by \cite{Juvela2008_CrA} where the $A_{\rm V}^{SCA}$ was below
the $A_{\rm V}^{NICER}$ values up to $\sim10^{\rm m}$ while the
situation was reversed at higher $A_{\rm V}$. On the southern side the
relation is $A_{\rm V}^{SCA}$=0.81+2.03$\times A_{\rm V}^{\rm sub-mm}$
with an even higher correlation coefficient, $r$=0.98.
The main difference compared to the northern side is the much steeper
slope of the relation. The northern and the southern fields were
calibrated separately, but a factor of two difference in the
calibration is very improbable. Furthermore, the surface brightness
values decrease to zero both at the southern and northern edges of the
maps and the surface brightness values were matched at the centre when
the fields were mosaiced together. Therefore, there is no room for an
over 60\% multiplicative error in the relative calibration. The
difference is an indication of a change in the local NIR radiation
field. Assuming that the grain properties are the same, and ignoring
the fact that the scattering is mostly in the forward direction, this
would suggest that the radiation field intensity is at least 50\%
higher on the southern side of the filament. This is qualitatively
consistent with the conclusions of \cite{Juvela2009_CrA} that were
based on the morphology of the NIR and the Spitzer dust emission maps.

The average ratios $A_{\rm V}^{SCA}$/$A_{\rm V}^{\rm sub-mm}$ and
$A_{\rm V}^{NICER}$/$A_{\rm V}^{\rm sub-mm}$ are $\sim 1.5$ or higher.
The interpretation of $A_{\rm V}^{SCA}$ is affected by the uncertainty
of the intensity of the interstellar radiation field (ISRF) but that
does not influence $A_{\rm V}^{NICER}$. It would thus appear that ISRF
is not very different from \citet{Mathis1983} values north of the
filament but are clearly elevated on the southern side. The $A_{\rm
V}^{\rm sub-mm}$ values appear to be too low by almost a factor of
two. Part of the difference could be caused the assumed scaling
between N(H$_2$) and $A_{\rm V}$, but it is clear that especially the
value of the dust opacity $\kappa$ has a large uncertainty. The
$A_{\rm V}^{\rm sub-mm}$ values were derived with an opacity of
$\kappa=$0.1\,cm$^2$/g\,($\nu$/1000\,GHz)$^{\beta}$ that is higher
than the value found in diffuse regions. \citet{Boulanger1996} derived
for high latitude clouds a value of $\kappa(850\,\mu{\rm
m})$=0.005\,cm$^2$\,g$^{-1}$ which is $\sim$40\% of the value given by
the previous formula. Thus, with the diffuse medium $\kappa$ the 
$A_{\rm V}^{\rm sub-mm}$ values would rise {\em above} the $A_{\rm
V}^{NICER}$ values. This suggests that the actual value of $\kappa$
should be somewhere in between. However, it is also known that the
colour temperature used in Eq.~\ref{eq:colden} overestimates the mass
averaged dust temperature and may lead to the underestimation of the
column densities \citep[see][]{Shetty2009a, Malinen2011,
YsardJuvela2012, Juvela2012_TB}. If this effect can be quantified, it
becomes possible to increase the lower limit the dust opacity. This is
possible only with modelling and we will return to this question in
Sect.~\ref{sect:modelling}.

\subsection{The NIR and sub-millimetre profiles of the filament}
\label{sect:profiles}

We measure the profile of the filament using the three tracers
available to us. The sub-millimetre emission is the only one that, at
least in principle, is capable of probing the column density
distribution across the whole filament. Good $A_{\rm V}^{NICER}$
estimates are missing for a small central part where no background
stars are visible (roughly the white area in
Fig.~\ref{fig:colden_comparison}a). The NIR surface brightness data
are reliable only in regions of less than $A_{\rm V}\sim 10^{\rm m}$.
Because we want to compare all the tracers, we concentrate on the
narrow region that is marked with dashed lines in
Fig.~\ref{fig:pretty_plot}d. We calculate the one-dimensional profile
along the main axis of that region, averaging the data in the
perpendicular direction.

\begin{figure}
\centering
\includegraphics[width=8cm]{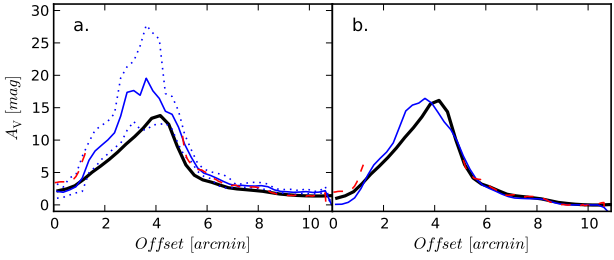}
\caption{
The column density profiles for the region marked with a dashed
rectangle in Fig.~\ref{fig:pretty_plot}d. The left hand frame shows
the $A_{\rm V}^{sub-mm}$ (thick black line), $A_{\rm V}^{NICER}$ (thin
blue line), and $A_{\rm V}^{SCA}$ (dashed red line, excluding the
central filament where the NIR emission becomes partly saturated) at
the original resolution (40$\arcsec$, 30$\arcsec$, and 10$\arcsec$,
respectively) without scaling. The x-axis is the offset along the
stripe, from south to north. 
The dotted lines are the profiles of the maximum and minimum $A_{\rm
V}^{\rm NICER}$ along the stripe.
The right hand frame shows the same data at a common
40$\arcsec$ resolution, with values set to zero at offset 10$\arcmin$
(additive correction). The $A_{\rm V}^{sub-mm}$ values have been
scaled by a factor 1.3 to match the other curves on the northern side
of the filament.
}
\label{fig:profiles}
\end{figure}

Figure~\ref{fig:profiles} shows the resulting column density profiles
that, like in Fig.~\ref{fig:colden_comparison}, are converted to units
of visual extinction. The first frames show the data at the original
resolution (cf Fig.~\ref{fig:colden_comparison}). There is an offset
between the maps. In the northern end the $A_{\rm V}^{\rm sub-mm}$
decreases below 0.5$^{\rm m}$ while the other maps are, based on the
large scale 2MASS extinction maps, were set at $\sim 2^{\rm m}$. This
is not likely to be caused by an uncertainty in the zero point of the
Herschel surface brightness data.  The missing $\sim 1.5^{\rm m}$
corresponds to a 250\,$\mu$m surface brightness of
$28$\,MJy\,sr$^{-1}$, this assuming a temperature of $T_{\rm d}=17$\,K
and a dust opacity of $\kappa=0.005 \times
(\lambda/850$\,$\mu$m)$^{-2}$\,cm$^2$\,g$^{-1}$ (see
Sect.~\ref{sect:compare_colden}). Such a surface brightness offset
would be comparable to the minimum signal found in the northern part
of examined area. 
Figure~\ref{fig:profiles} again shows the discrepancy in the $A_{\rm
V}^{\rm SCA}$ estimates. On the northern side these follow closely the
shape of the $A_{\rm V}^{\rm NICER}$ curve but they are significantly
higher in the southern end. In the plot we have not included the
$A_{\rm V}^{\rm SCA}$ values for the central filament, $A_{\rm V}^{\rm
SCA}>10^{\rm m}$, because of their large uncertainty near the region
where the surface brightness saturates.

The frame $b$ of Fig.~\ref{fig:profiles} shows the same data convolved
to a common resolution of 40$\arcsec$. The value at the offset of
$10\arcmin$ has been subtracted, and the $A_{\rm V}^{\rm sub-mm}$
values have been scaled by a factor of 1.3. With this scaling, the
column density estimates are very consistent on the northern slope.
For $A_{\rm V}^{\rm NICER}$ and the scaled version of $A_{\rm V}^{\rm
sub-mm}$ the match is good up to $A_{\rm V}=15^{\rm m}$. However, the
peak of $A_{\rm V}^{\rm NICER}$ is almost one minute of arc south of
the $A_{\rm V}^{\rm sub-mm}$ maximum. The $A_{\rm V}^{\rm sub-mm}$
profile is strongly skewed towards the north and rough agreement is
found again in the south end of the map.  If we trust the NICER map,
both $A_{\rm V}^{\rm SCA}$ and $A_{\rm V}^{\rm sub-mm}$ show
deviations on the south side of the filament and qualitatively both
features could be affected by the asymmetry of the radiation field.
Figure~\ref{fig:profiles}a shows as dotted lines the minimum and
maximum $A_{\rm V}^{\rm NICER}$ along the stripe for which the other
curves show the average value. These indicate that there is a strong
$A_{\rm V}$ gradient perpendicular to the stripe and parallel to the
filament, especially close to the high column density clump. This could
cause significant bias in $A_{\rm V}^{\rm NICER}$
\citep[see][]{Juvela2008_CrA} but cannot explain all the differences
between $A_{\rm V}^{\rm NICER}$ and $A_{\rm V}^{\rm sub-mm}$, neither
in the levels nor in the profiles.

\subsection{Mid-infrared profiles} \label{sect:coreshine}

\begin{figure}
\centering
\includegraphics[width=8cm]{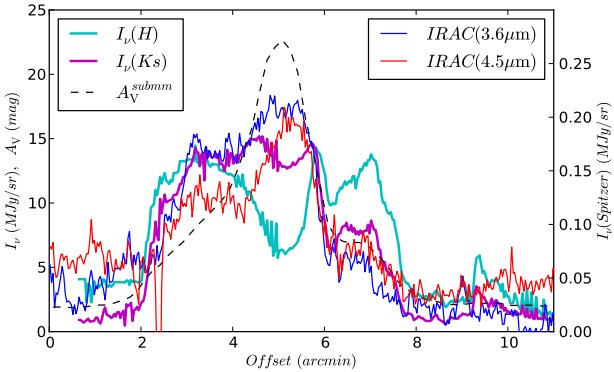}
\caption{
The mid-infrared profiles of the Corona Australis filament. The
offsets are the same as in Fig.~\ref{fig:profiles}. In addition to the
Spitzer 3.6\,$\mu$m and 4.5\,$\mu$m data (blue and lines,
respectively, and the right hand y-axis) the figure includes the
profiles for the scattered light in the $H$ and $Ks$ bands (cyan and
magenta lines, left hand axis) and the $A_{\rm V}$ derived from the
Herschel observations (black dashed line, left axis).
}
\label{fig:cs}
\end{figure}

Although we will not attempt to model the mid-infrared data of
the Corona Australis cloud (Sect.~\ref{sect:modelling}) it is also useful
to examine the Spitzer IRAC and the WISE satellite observations
of the filament. Figure~\ref{fig:cs} compares the 3.6\,$\mu$m and
4.5\,$\mu$m Spitzer IRAC profiles with the scattered light and the
extinction derived from the Herschel data.

The mid-infrared profiles show some resemblance to the NIR bands but
follow more closely the column density profile of the filament as
derived from the sub-millimetre observations. In particular, the
4.5\,$\mu$m emission peak coincides with the location of the column
density maximum derived from the Herschel data. Compared to the
estimated column density, both the 3.6\,$\mu$m and 4.5\,$\mu$m data
show significantly higher intensity levels on the south side. We
compared the Spitzer and WISE data at the wavelengths of 3.4\,$\mu$m
and 4.6\,$\mu$m. The WISE measurements are in complete agreement with
the Spitzer data at the corresponding wavelengths except for the
3.4\,$\mu$m intensity that is slightly lower in WISE at the location
column density peak ($\sim$0.18\,MJy/sr vs. 0.21\,MJy/sr) so that the
intensity remains almost flat between offsets
3.0$\arcmin$-5.5$\arcmin$. Further outside the filament, the Spitzer
ratio 3.6\,$\mu$m/4.5\,$\mu$m and the WISE ratio
3.4\,$\mu$m/4.6\,$\mu$m show some increase. This could be caused by
the effect the optically thick filament has on the local radiation
field but the quantification of these effects would require separate
modelling.

The mid-infrared data have revealed in many dense cloud cores the
presence of the ``coreshine'' phenomenon where the emission at
$\sim$3.5\,$\mu$m becomes brighter relative to the emission at
$\sim$4.5\,$\mu$m \citep{Steinacker2010, Pagani2010}. The changes are
interpreted as a sign of an increase of dust grain sizes that leads to
enhanced light scattering at wavelengths beyond 3\,$\mu$m. The first
detection of coreshine was made in the cloud LDN\,183 using Spitzer
data \citep{Steinacker2010}. Further detections have subsequently
been made with WISE observations \citep{PaperIII} but LDN\,183
remains the best example of the phenomenon.
Figure~\ref{fig:cs} shows significant increase of the
3.6\,$\mu$m/4.5\,$\mu$m ratio but only south of the column density
peak. This is thus probably caused by a change of the radiation field
rather than by grain growth. Regarding the overall asymmetry with
respect to the filament centre, the 3.5\,$\mu$m profile is rather
similar to that of the $K$ band.

\begin{table}
\caption{The radiative transfer models.}
\begin{tabular}{llll}
\hline \hline
Radiation field           &   Dust model   &   Figures         & Notes   \\
\hline
isotropic, $\chi=2$       &   OH           &   Fig.~\ref{fig:EI_chi}   &         \\
isotropic, $\chi=3$       &   OH           &   Fig.~\ref{fig:EI_chi}   &         \\
isotropic, $\chi=4$       &   MWD          &   Figs.~\ref{fig:EI4.0}, \ref{fig:sca_iso}   &  \\
isotropic, $\chi=4$       &   OH           &   Figs.~\ref{fig:EI4.0}, \ref{fig:EI4.0_oh_colden}, \ref{fig:sca_iso}, \ref{fig:laboca_EI4.0_oh} &  \\
anisotropic, $\chi=3+1^{a}$ &   OH         &   Figs.~\ref{fig:EA_oh_3.0_1.0}, \ref{fig:EA4.0_oh_colden}, \ref{fig:sca_aniso} &  \\
anisotropic, $\chi=3+1^{a}$ &   MWD        &   Fig.~\ref{fig:sca_aniso} &  \\
isotropic, $\chi=4$       &   MWD          &   Fig.~\ref{fig:sca_n2.0}  &  $n \times 2^{b}$  \\
anisotropic, $\chi=3+1$   &   MWD          &   Fig.~\ref{fig:sca_n2.0}  &  $n \times 2^{b}$  \\
\hline
\end{tabular}

$^{a}$ Ratio 3:1 between isotropic background and additional radiation from the 
south-eastern direction.

$^{b}$ Densities twice the values obtained from the modelling of sub-millimetre emission.
\end{table}

\section{Radiative transfer models} \label{sect:modelling}

\subsection{Model of the sub-millimetre emission}
\label{sect:model-submm}

We construct a three-dimensional model to explain the Herschel
observations of 160\,$\mu$m, 250\,$\mu$m, 350\,$\mu$m, 500\,$\mu$m. In
the present paper we examine only the main effects, including the
possibly anisotropic radiation field, using two dust models. The first
model corresponds to $R_{\rm V}$=5.5 and is described in
\cite{Draine2003}\footnote{The data files describing the dust
properties are available at {\tt http://www.astro.princeton.edu/
draine/dust/}}. In the following, this dust model is called MWD. As a
point of comparison, we use the \cite{Ossenkopf1994} dust model (in
the following, the model OH) for coagulated grains with thin ice
mantles that have accreted in $10^5$ years at a density of
$10^5$\,cm$^{-3}$. At wavelengths below 1\,$\mu$m we adopt the short
wavelength extension discussed in \cite{Stamatellos2003}. The dust
models differ with respect to the sub-millimetre spectral index
$\beta$ that, measured between 250\,$\mu$m and 500\,$\mu$m is $\sim
2.1$ for the MWD dust model and 1.76 for the OH model. The optical
depth ratios for extinction $\tau(0.55\,\mu{\rm m})/\tau(350\,\mu{\rm
m})$ are 1760 and 2080 for MWD and OH, respectively. Detailed studies
of the possible variations of the dust properties are deferred to a
later paper. Here the two models are used only to test the sensitivity
of the results to the actual dust properties. 
The dust emission is calculated with our radiative transfer
program \citep{Juvela2003_CRT, Juvela2005_CRT}. The program uses Monte
Carlo simulation to determine the radiation field intensity at each
position within the model cloud. This information is used to determine
the distribution of dust temperatures. The line-of-sight integration
of the radiative transfer equation then results in synthetic surface
brightness maps that are calculated for each of the observed
wavelengths.

To avoid edge effects near the boundaries of the area covered by the
NIR data, we include in the model a larger area than that shown in
Fig.~\ref{fig:pretty_plot}. The modelled area is $10\arcmin \times
10\arcmin$, centred at coordinates $\alpha_{\rm 2000}$=19$^{\rm
h}$0$^{\rm m}$54.3$^{\rm s}$, $\delta_{\rm
2000}$=-36$\degr$55$\arcmin$22$\arcsec$. 
The model will be adjusted to reproduce the observed 350\,$\mu$m data.
This is accomplished by having as free parameters the column densities
corresponding to each map pixel. The optimisation will determine the
column density for each line-of-sight and thus the mass distribution
in the plane of the sky. In principle, the solution should be unique for
any combination of the external radiation field and dust properties.
However, the line-of-sight density distribution must be fixed for
this modelling. According to Fig.~\ref{fig:profiles} in the plane of
the sky the FWHM of the column density distribution is
$\sim2.5\arcmin$ perpendicular to the filament. For the assumed
distance of 130\,pc \citep{Marraco1981} this corresponds to a filament
width of $\sim 0.1$\,pc. 
We assume that the filament is approximately cylindrical. Thus,
in the model the density distribution along the line-of-sight is set
to be Gaussian with FWHM equal to 0.09\,pc. The sensitivity of the
results on this assumption is discussed further in
Sect.~\ref{sect:modelling}.
The model is modelled by a Cartesian grid of 60$^3$ cells with
the cell size corresponding to 10$\arcsec$. The surface brightness
maps are calculated towards one principal axis, the simulated map
consisting of $60 \times 60$ pixels. When the model is compared
with observations, the model results are convolved to the resolution
of the corresponding observation. In particular, at 350\,$\mu$m the
observations have a resolution of $\sim25 \arcsec$ while the model
discretisation corresponds to an angular resolution of 10$\arcsec$.

We start by assuming that the filament is illuminated by an isotropic
ISRF with the spectrum given by \citet{Mathis1983}. The solution is
obtained iteratively, calculating the model prediction of the
350\,$\mu$m surface brightness and scaling independently the column
density corresponding to each of the $60\times 60$ map pixels.  The
densities of cells corresponding to a single map pixel are scaled with
the same number, thus preserving the original shape of the
line-of-sight density profile.
The first realisation is that the CrA cannot be modelled using the
standard value of the ISRF. The column densities increase without
limit but the models never reach the surface brightness observed at
the centre of the filament. There are two solutions to this problem.
Either the sub-millimetre opacity must be increased by at least 
factor of $\sim$2 (or more considering the associated decrease in dust
temperatures). The other alternative is to increase the radiation
field.

\begin{figure}
\centering
\includegraphics[width=8.3cm]{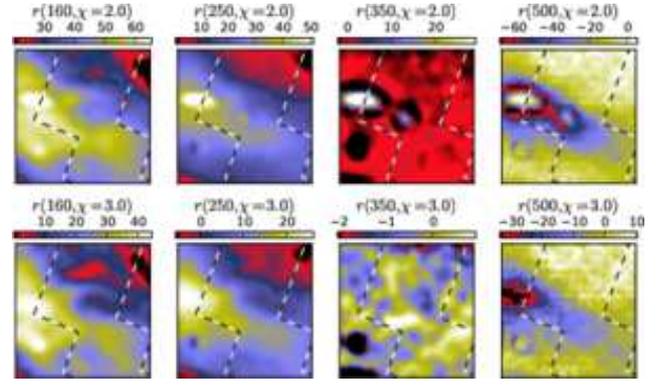}
\caption{
The fit residuals for models with the OH dust and the radiation field
intensity scaled by factors $\chi$=2 and 3. The values are relative
errors $I_{\rm Obs}-I_{\rm Mod}$ in per cents of the observed
surface brightness $I_{\rm Obs}$.
The area is this and the subsequent plots of model results is
$10\arcmin \times 10\arcmin$ and the plots are centred at coordinates
$\alpha_{\rm 2000}$=19$^{\rm h}$0$^{\rm m}$54.3$^{\rm s}$,
$\delta_{\rm 2000}$=-36$\degr$55$\arcmin$22$\arcsec$. 
}
\label{fig:EI_chi}
\end{figure}

Figure~\ref{fig:EI_chi} shows the fit residuals in the case of the OH
dust model and the scaling of the radiation field by factors $\chi$=2
and 3. With $\chi=2.0$ the surface brightness never reaches the
observed values, the error at 350\,$\mu$m remaining above 50\%. With
$\chi=3.0$ the 350\,$\mu$m can be fitted but only when the peak
$A_{\rm V}$ values are above 100$^{\rm m}$ over a significant area of
the map. Apart from being incompatible with the $A_{\rm V}^{\rm
NICER}$ data, the model fails to produce the correct shape of the
emission spectrum. For example, the 160\,$\mu$m surface brightness is
too low typically by more than 20\%.

\begin{figure}
\centering
\includegraphics[width=8.3cm]{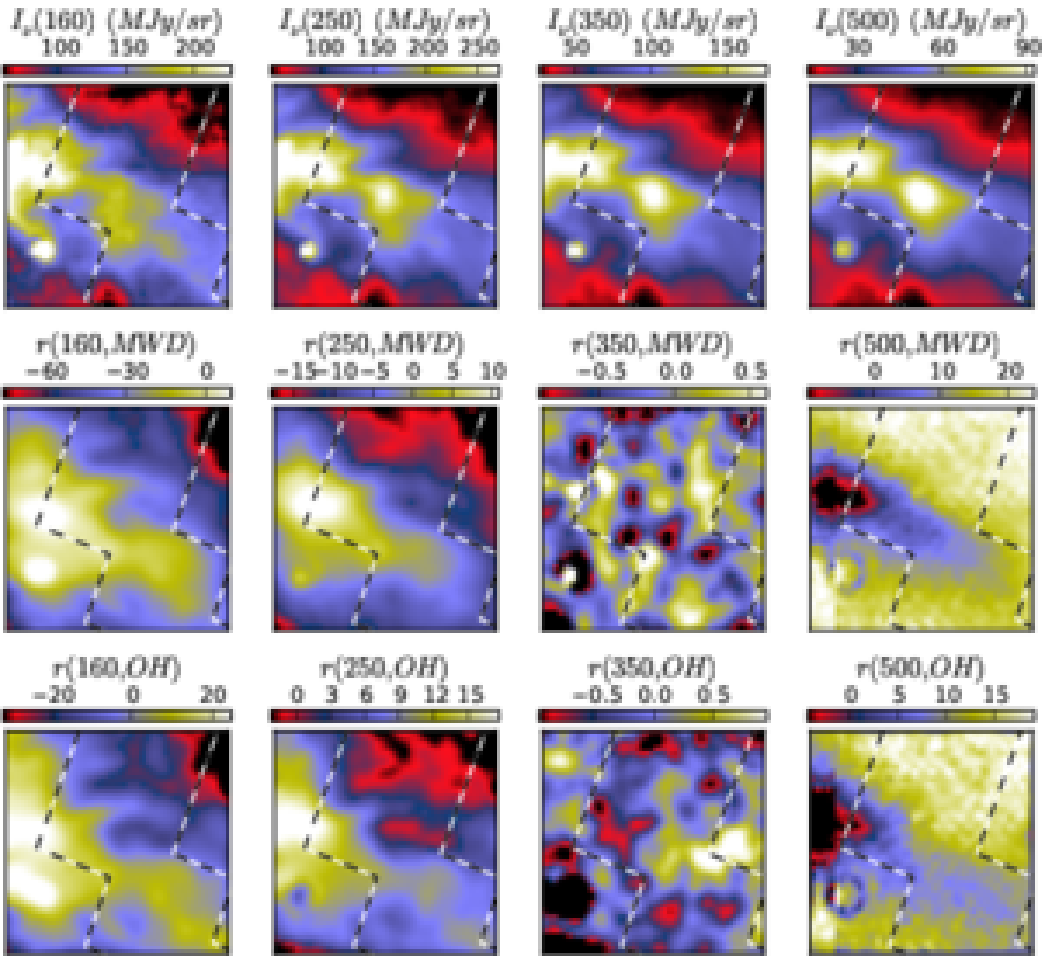}
\caption{
A model fit to the Herschel surface brightness data with an isotropic
radiation field four times the ISRF of \cite{Mathis1983}.
The first row shows the original observations. The other rows show the
errors, ($I_{\rm Obs}-I_{\rm Mod}$)/$I_{\rm Obs}$,
in units of per cent. The second row corresponds to the MWD dust
model \citep{Draine2003}, and the bottom row to the OH dust model
\citep{Ossenkopf1994}.
To exclude the regions affected by boundary effects, each frame shows
the central $8.3\arcmin \times 8.3\arcmin$ area of the whole model
that covered an area of $10\arcmin \times 10\arcmin$. To guide the
eye, the dashed lines show the outline of the area covered by the NIR
observations.
}
\label{fig:EI4.0}
\end{figure}

More satisfactory solutions are found by scaling the ISRF values by a
factor of $\chi=$4.  Figure~\ref{fig:EI4.0} shows the observed surface
brightness maps and the fit residuals for both the MWD and OH dust
models. Because the model is fitted only based on the 350\,$\mu$m
intensity, and the product of the radiation field and column density
is capable of producing high enough surface brightness, the rms error
at this wavelength is now below 1\%. 

Concentrating on the central area, the fit is better with the OH
model. Apart from the noise affecting the low column density part, the
relative errors are $\sim$10\% at both 250\,$\mu$m and 500\,$\mu$m.
There is no strong overall bias even at 160\,$\mu$m. This indicates
that the average shape of the SED is well reproduced. There are,
however, some systematic effects. The observed level of the
500\,$\mu$m emission is well reproduced in the filament but is
overestimated is the lower density regions. The error is small,
$\sim$10\%, and could be associated with the uncertainty of the
intensity zero point of the observations. 
A stronger effect is seen at the location of the eastern clump, near
the boundary of the area shown. There the observed 500\,$\mu$m surface
brightness is up to $\sim$15\% lower than the model prediction. The
percentage error is not large but it is very clear in the map and
significant considering that the nearby wavelength of 350\,$\mu$m is
perfectly fitted. In the same area the model fails to produce
sufficiently high surface brightness at short wavelengths, the error
increasing to $\sim$30\% at 160\,$\mu$m. One possible interpretation
is that the actual radiation field is higher in this area. By
underestimating the heating, the model would require a column
density that is too high to explain the 350\,$\mu$m map (hence the high 500\,$\mu$m
signal) but does not include enough warm dust to reproduce the shorter
wavelengths. 
The 160\,$\mu$m (and 250\,$\mu$m) residuals show a general gradient
towards south-east that can indicate an increase of the radiation
field intensity. This would be consistent with the higher level of NIR
scattered light seen in that direction (see
Sect.~\ref{sect:profiles}).
An alternative explanation for the behaviour of the eastern clump
involves the dust opacity spectral index. A higher $\beta$ value might
be needed to correct the ratio between the 350\,$\mu$m and 500\,$\mu$m
data that are relatively insensitive to variations in the dust
temperature. 

The error maps of the MWD model look qualitatively similar to the OH
results but the $\chi^2$ value (considering maps other than the
350\,$\mu$m) is worse by more than a factor of two. Outside the
filament the 160\,$\mu$m and the 250\,$\mu$m signals are mainly
overestimated and thus the emission appears too warm. However, the
250--500\,$\mu$m spectral index of MWD is 2.1, i.e., higher than the
value $\beta$=1.76 of the OH model.Therefore, compared to the OH dust
model, the best fit would also be expected to correspond to a somewhat
lower value of the radiation field.
In the eastern clump the 500\,$\mu$m emission is again overestimated
and the $160\,\mu$m emission underestimated but, possibly because of
the different $\beta$ values, the errors are smaller. There is still a
gradient towards south-east consistent with an increase in of the dust
temperature. However, as already noted, the sign of the 160\,$\mu$m
errors has changed compared to the OH model, and the MWD model would produce
insufficient intensity at the southern side of the filament.

\begin{figure}
\centering
\includegraphics[width=8.3cm]{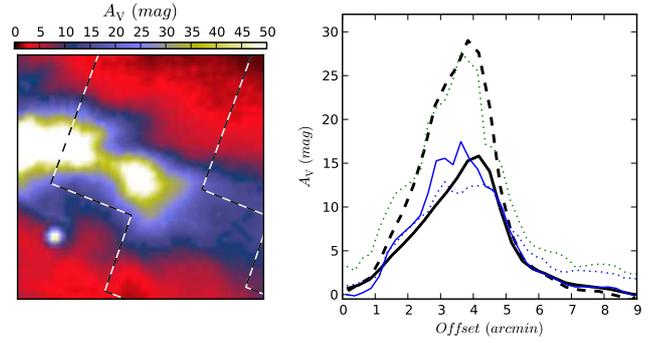}
\caption{
The visual extinction in the model with OH dust and an isotropic
radiation field with the intensity four times the standard value
\citep{Mathis1983}. The left hand frame shows the $A_{\rm V}$ map 
from the model smoothed to a resolution of 18$\arcsec$. The right
hand frame shows the $A_{\rm V}$ profile of the model
(dashed line) compared to the observed profiles of $A_{\rm
V}^{\rm sub-mm}$ (thick solid line) and $A_{\rm V}^{\rm NICER}$ (thin
blue line) along the selected stripe. The latter two are the same as
in Fig.~\ref{fig:profiles}b but are shown at their original
resolution. In the plot the $A_{\rm V}^{\rm sub-mm}$ has been scaled
by a factor of 1.3. The dotted lines are the profiles of the maximum
and minimum $A_{\rm V}^{\rm NICER}$ along the stripe.
}
\label{fig:EI4.0_oh_colden}
\end{figure}

We show in Fig.~\ref{fig:EI4.0_oh_colden} the $A_{\rm V}$ map and the
filament profile for the OH model where the \citet{Mathis1983}
radiation field was scaled by a factor of four. Because of the
isotropy of the radiation field, the shape of the recovered profile is
similar to that of $A_{\rm V}^{\rm sub-mm}$. However, the predicted
$A_{\rm V}$ values are not just higher than the previous $A_{\rm
V}^{\rm sub-mm}$ estimates, which is possible for a number of reasons,
but also higher than the measurements of $A_{\rm V}^{\rm NICER}$. This
is an indication that either the assumption of the NIR extinction
curve (used in the conversion of NIR colour excesses to $A_{\rm V}$)
is incorrect or, more likely, the ratio $\tau(0.55\,\mu{\rm
m})/\tau(350\,\mu{\rm m})$ used in the modelling is incorrect. An
increase in the sub-millimetre opacity would decrease the model
$A_{\rm V}$. If this modification were restricted to the centre of the
filament, the radiation field would not have to be changed and the
correct shape of the SED would still be recovered at low and
intermediate column densities. In the same figure we also show again
the minimum and maximum $A_{\rm V}^{\rm NICER}$ profiles along the
selected stripe (dotted lines). The shape and magnitude of $A_{\rm
V}^{\rm sub-mm}$ in again seen to be in agreement with the maximum
$A_{\rm V}^{\rm NICER}$ profile rather than the average profile.

The previous study of the NIR observations and Spitzer FIR data
suggested an anisotropy with a stronger radiation field in the south
\citep{Juvela2009_CrA}. Figure~\ref{fig:profiles} lead to a similar
conclusion. Because the Galactic plane is located north of Corona
Australis cloud, there must be a more local cause for the asymmetry.
The star S CrA is located on the south side of the filament,
immediately east of the area covered by our NIR data. The Herbig Ae/Be
star R Corona Australis is farther in the east \citep{Gray2003}. The
stars appear to affect the dust emission at least in the eastern part
of our maps (as seen in Fig.~\ref{fig:EI_chi}). The precise spectral
type of S CrA is not known (the main component S CrA A was classified
as G5Ve by \citet{Carmona2007}). Based on the observed K band
magnitude of S CrA, $m_{\rm K}=6.1^{\rm m}$, the NIR intensity
produced by S CrA is comparable to the general ISRF up to distances
corresponding to several arc minutes. R CrA may have similar
contribution to the NIR radiation field although, because of large
intervening extinction, its effect on dust heating is probably
smaller.
The other bright stars in the region, HD\,176269 and HD\,176270, are
near the south end of the area covered by our NIR observations (see
Fig.~\ref{fig:IRAC}). The distance estimates of these B9V stars are
uncertain but they may both be associated with the cloud
\citep[see][]{Peterson2011}. 

We examine the situation by illuminating the model cloud with an
isotropic ISRF component and adding a second radiation source in the
southern direction, at a position angle of 20 degrees from south to
east. The source is assumed to be at a large distance. Both radiation
field components are assumed to have the same spectral shape as the
normal ISRF and their combined energy input is kept the same as in
Fig.~\ref{fig:EI4.0_oh_colden}, four times that of the normal ISRF.

\begin{figure}
\centering
\includegraphics[width=8.3cm]{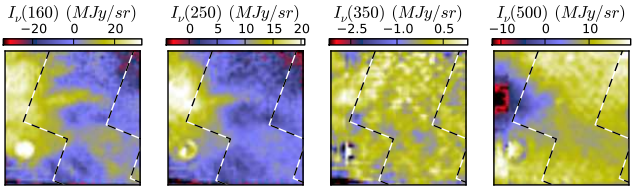}
\caption{
The relative errors (\%) in the predicted surface brightness in the case of
an anisotropic radiation field and the OH dust model.
}
\label{fig:EA_oh_3.0_1.0}
\end{figure}

\begin{figure}
\centering
\includegraphics[width=6.5cm]{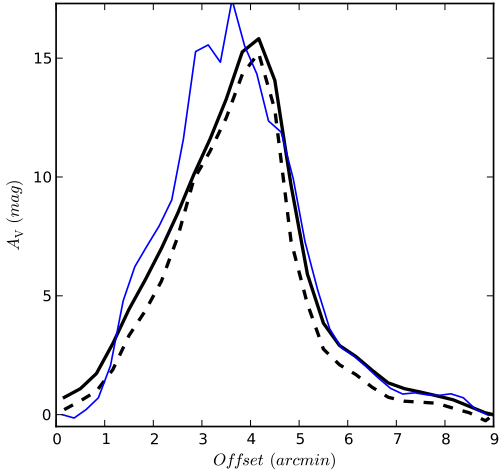}
\caption{
The $A_{\rm V}$ profile of the filament as determined from the 
radiative transfer model with an anisotropic radiation field (dashed
lines, scaled with 0.5). The profile is compared with the $A_{\rm
V}^{\rm NICER}$ (blue solid line) and the $A_{\rm V}^{\rm sub-mm}$
profile (solid black line, scaled with 1.3). 
}
\label{fig:EA4.0_oh_colden}
\end{figure}

When the southern source stands for 25\% of the total radiation field
energy, the north-south gradients in the residuals of the 160\,$\mu$m
and 500\,$\mu$m fits disappear (Fig.~\ref{fig:EA_oh_3.0_1.0}).  Thus
the north-south radiation field anisotropy is now correct as far as
can be concluded from the FIR and sub-mm emission. The result
assumes that the additional radiation is coming directly from the SE
direction. If the direction for the incoming radiation is not
perpendicular to the line-of-sight (e.g., more on the front side of
the cloud), the actual anisotropy of the radiation field may be
larger. 
Because the dust is now warmer on the southern side of the filament
the maximum of the column density should move towards north thus
increasing the discrepancy with $A_{\rm V}^{\rm NICER}$ that peaks
south of $A_{\rm V}^{\rm sub-mm}$. However, as shown in
Fig.~\ref{fig:EA4.0_oh_colden}, the effect on the shape of the model
column density profile is not significant. The gradient in the
east-west direction still remains unexplained by the model. More
specifically, the clump near the east boundary of the NIR map shows,
compared to our model, an excess at 250\,$\mu$m and low surface
brightness at 500\,$\mu$m. As already noted, this can be a sign of a
further temperature gradient (natural considering the vicinity of the
R CrA region) possibly combined with an increase in the dust spectral
index.

If the ISRF level is further increased by 20\%, the column densities
decrease but the $\chi^2$ values are higher by more than a factor of
two, mainly because the 160\,$\mu$m intensities are overestimated
throughout the field. This indicates that, apart from the inherent
uncertainties involved in the modelling, the total intensity of the
radiation field is well constrained. The models cannot produce
sufficient surface brightness with lower level of heating and with
increased ISRF the spectrum becomes incorrect. However, if the dust
opacity in the cloud centre increases by a factor of a few, the lower
limit of the ISRF could be relaxed correspondingly.

\subsection{Modelling of the NIR scattered light}

We calculate the NIR surface brightness maps for the previous models
where the radiation field corresponded to four times the normal ISRF
\citep{Mathis1983} and the field is either isotropic or anisotropic.
In the models of the sub-millimetre dust emission, the density
distributions are similar for the MWD and the OH dust, the column
densities being only $\sim20$\% higher for the MWD. Therefore, in the
following we consider only the density distributions derived from the
fits of the Herschel data with the OH dust model. With this fixed
density distribution, the scattered light can be calculated using both
the MWD and the OH dusts. The OH model does not specify the scattering
function (i.e., the directional distribution of scattered photons) and
we use the same scattering function as in the case of MWD. 

\begin{figure}
\centering
\includegraphics[width=8.3cm]{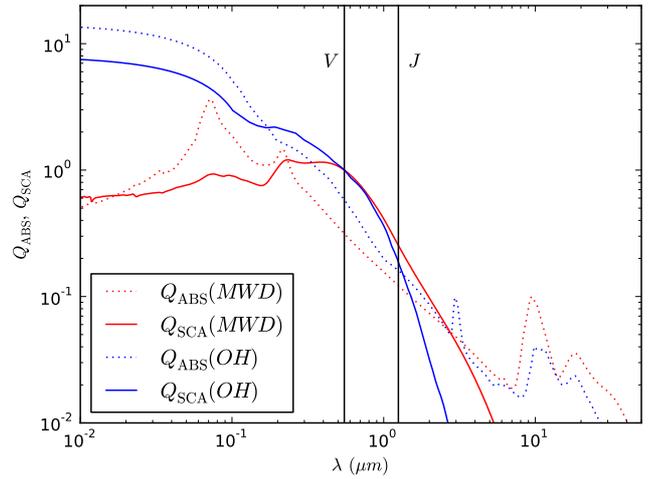}
\caption{
The scattering and absorption efficiencies for the two dust models.
The solid curves show the scattering efficiencies $Q_{\rm Sca}$ for
the MWD dust (solid red line) and the OH dust (solid blue line) that
have been normalised to a value of 1.0 in the $V$ band (0.55\,$\mu$m).
The dotted lines are the corresponding efficiencies for absorption,
$Q_{\rm Abs}$, that have been scaled with the same factors as the
$Q_{\rm Sca}$ values.
}
\label{fig:sca}
\end{figure}

The differences between the optical and near-infrared properties of
the two dust models are important as can be seen in
Fig.~\ref{fig:sca}. For MWD the scattering is still dominant in the
$J$ band while for the OH model the efficiencies $Q_{\rm Sca}$ and
$Q_{\rm Abs}$ are almost equal. In $J$ band the albedos are 0.68
and 0.58 for MWD and OH, respectively. In case of multiple scattering,
the effect on the surface brightness can be expected to be noticeable.

\begin{figure}
\centering
\includegraphics[width=8.3cm]{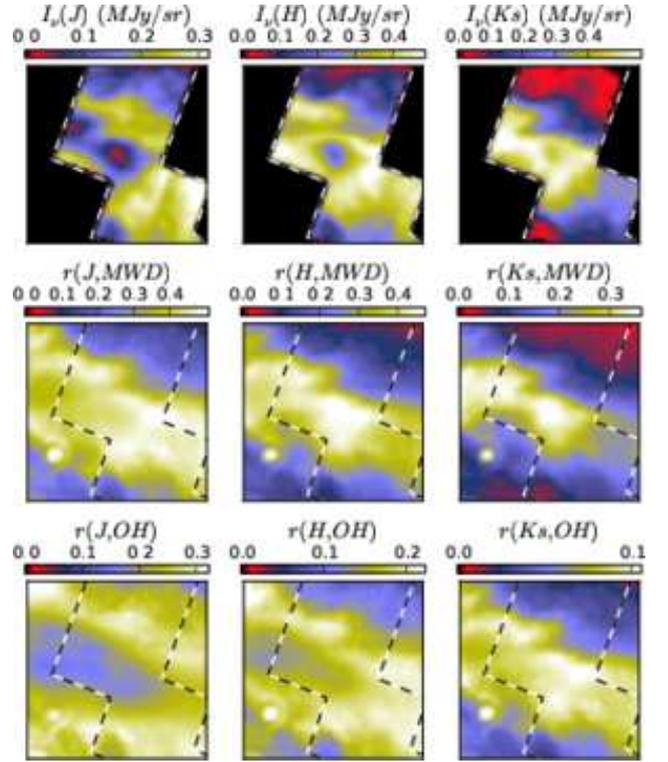}
\caption{
The observed NIR surface brightness (first row) and the surface
brightness predicted for the models based on the Herschel
sub-millimetre data (OH dust, $\chi=$4.0). The second and the third
rows correspond to calculations of the scattered light with the MWD and
the OH dust models, respectively, in case of isotropic illumination.
}
\label{fig:sca_iso}
\end{figure}

The surface brightness images are shown in Fig.~\ref{fig:sca_iso}. For
the MWD dust, the predicted surface brightness levels are of the right
order of magnitude. The $H$ band levels are very close to the right
value, the $J$ band values are $\sim$30\% higher than the observed
values and the $K$ band values are lower by a similar amount. This
could point to a NIR spectrum of the ISRF that is redder than in the
\cite{Mathis1983} model.  
\cite{Juvela2008_CrA} obtained similar results noting that in
particular the $H$ and $K$ band values were twice as high as expected
in the case of the normal ISRF. In the present study, the modelling
also takes into account the effect that the optically thick filament
has on the radiation field. This shadowing is probably the main reason
why the modelling prefers higher values for the intensity of the
external radiation field.

The NIR calculations fail to reproduce the surface brightness dip that
is evident in the observations. If the modelling of the dust emission
underestimates the column density at the centre of the filament, this
could also partly explain the difference in the NIR colours. Because
the NIR scattering takes place is mostly in the forward direction, an
increase in the column density would make the spectrum of the
scattered light redder. Higher column densities would appear to be
incompatible with the $A_{\rm V}^{\rm NICER}$ data but, because no
stars are seen through the centre of the filament \citep[see Fig. 7
in][]{Juvela2008_CrA}, this cannot be excluded on the basis of $A_{\rm
V}^{\rm NICER}$ alone. The present model is based on the modelling of
dust emission and already has a peak extinction that is twice the
maximum of the $A_{\rm V}^{\rm NICER}$ map. If the radiation field
intensity is decreased by a factor of two, the column density could
increase almost without limit and there would be no constraints on the
maximum column density. This is excluded only by the fact that a lower
intensity of the radiation field would alter the SED of the dust
emission in a way that is incompatible with the observations.

The OH dust model (Fig.~\ref{fig:sca_iso}, bottom row) reproduces the
morphology of the NIR data better than the MWD dust model. The dip in
the $J$ band is now noticeable at the location of the dense clump. The
depth of the depression is about half of the peak values around the
central clump. This is still far from the almost complete absence of
scattered light in the observations but is clearly a step in the right
direction. A small depression is seen even in the $Ks$ band where the
MWD dust produced a clear peak. The maximum $J$ band intensity is
again slightly higher than the observed values. On the other hand, the
$Ks$ band signal is very low, only one quarter of the observed.

\begin{figure}
\centering
\includegraphics[width=8.3cm]{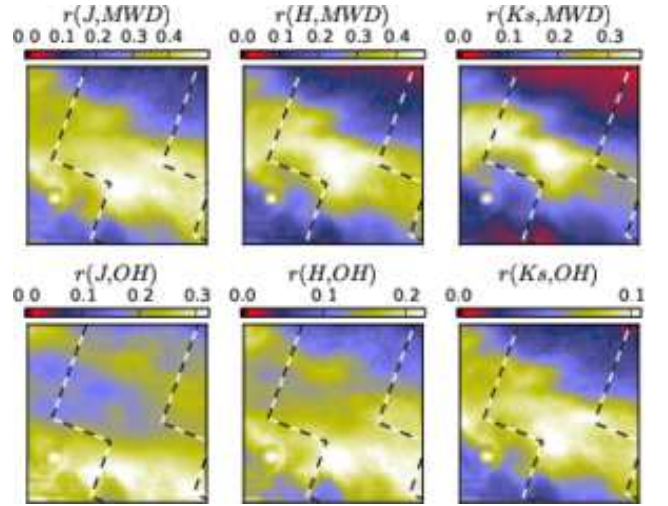}
\caption{
The NIR surface brightness for the models with anisotropic radiation
field. The first and the second row correspond to calculations of the
scattered light with the MWD and the OH dust models, respectively.
}
\label{fig:sca_aniso}
\end{figure}

We show in Fig.~\ref{fig:sca_aniso} similar calculations for the model
of Fig.~\ref{fig:EA_oh_3.0_1.0}. The scattered light is calculated
consistently assuming that, in addition to an isotropic radiation
field component, 25\% of the total radiation comes from the southern
direction. The resulting surface brightness asymmetry between the
south and north side of the filament is already slightly too large.
Although the ratio of 3:1 between the isotropic and the anisotropic
radiation field components was appropriate for the distribution of
dust emission, the NIR surface brightness at least precludes larger
degree of anisotropy, assuming that the spectrum of the anisotropic
source is the same as for the isotropic component.

\begin{figure}
\centering
\includegraphics[width=8.3cm]{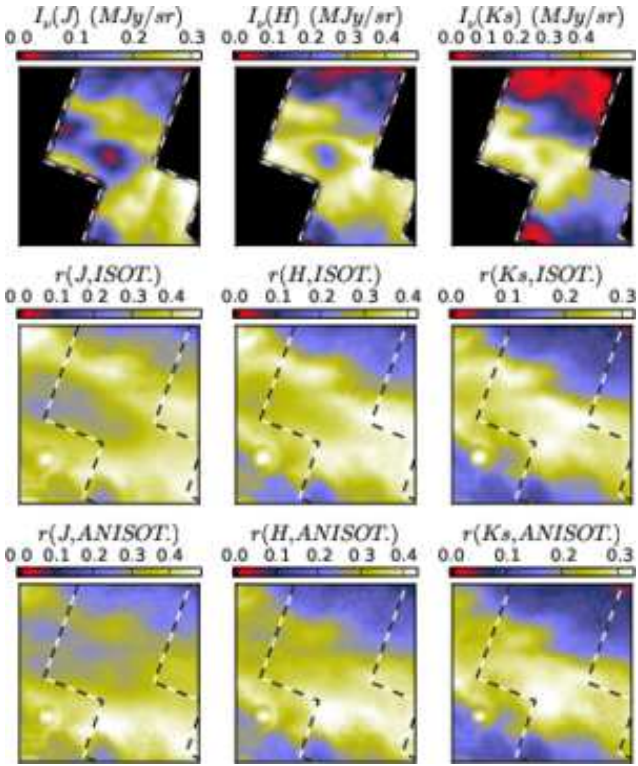}
\caption{
The observed NIR surface brightness (first row) and the surface
brightness predicted by models with the densities twice the values of
those fitting the sub-millimetre data. The second and the third row
correspond to the cases of isotropic and anisotropic radiation field,
respectively.
}
\label{fig:sca_n2.0}
\end{figure}

The major shortcoming of the models is the absence of a sufficiently
strong surface brightness dip at the location of the main clump. This
could be caused either by the column densities being underestimated
(unlikely given the extinction data) or by some modification of the
dust properties.
To check the first alternative and to check that the previous
differences between the OH and MWD models were not just the effect of
different opacities (instead of the albedo), we recalculated the MWD
model after scaling its densities by a factor of two. This results
only a very minor improvement (see Fig.~\ref{fig:sca_n2.0}) although
the maximum of the $A_{\rm V}$ map is already well over $100^{\rm m}$.
The effect on the morphology of the NIR maps is smaller than the
difference between the two dust models.

\begin{figure}
\centering
\includegraphics[width=8.3cm]{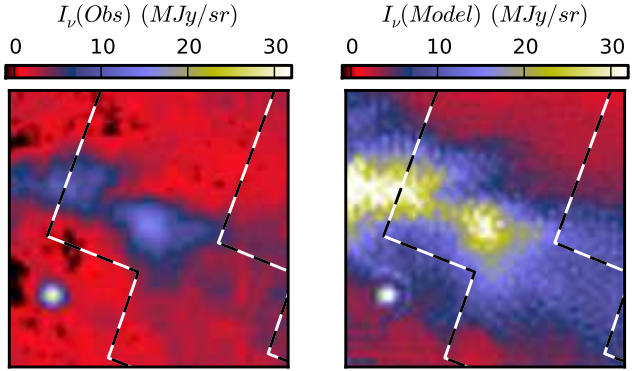}
\caption{
Comparison of the observed 870\,$\mu$m surface brightness and
the prediction from the model with $\chi=4.0$ with the OH dust.
}
\label{fig:laboca_EI4.0_oh}
\end{figure}

The models predict higher 870\,$\mu$m intensities than those
observed with LABOCA (see Fig.~\ref{fig:laboca_EI4.0_oh}). However, a
large fraction of the difference can be explained by the spatial
filtering that removes large scale structure from the LABOCA map.
In the NIR observations the $J$ band surface brightness goes
almost to zero at the centre of the main clump. This is difficult to
explain because some scattered light is always observed from the outer
cloud layers. Thus, the explanation may require a very special
geometry (e.g., a smaller line-of-sight extent of the cloud at the
location of the clump) and/or further modifications to the dust models
as suggested by the difference between the MWD and OH model results. 
Although the NIR signal saturates by $A_{\rm V}\sim 20^{\rm m}$, the
data still set very strong constraints both on the cloud structure and
the dust properties.

\subsection{Sensitivity to the cloud shape}

In the modelling the column density structure is determined directly
by the observations. For fixed dust properties and a fixed external
radiation field the result should be unique. However, there are still
factors connected with the density distribution that can affect the
results. The actual three-dimensional shape of the cloud is unknown.
Although the morphology suggests the presence of a single cylindrical
filament, the cloud could still be flattened or elongated along the
line-of-sight. Some effects can also result from small scale
inhomogeneities that allow short wavelength radiation to penetrate
deeper into the cloud. We tested the sensitivity of the results to
these factors using the model presented in
Fig.~\ref{fig:EA_oh_3.0_1.0}. 

The clumpiness was implemented by multiplying the density of each cell
in the model by a random number uniformly distributed between zero and
one. This results in a fair amount of inhomogeneity considering that
the cloud radius is only $\sim$30 cells. All densities were then
rescaled so that the model again reproduced the observed 350\,$\mu$m
surface brightness. Compared to the original model with a smooth
density distribution, this resulted in less than 10\% changes in the
column density along individual lines-of-sight. More importantly, this
level of inhomogeneity did not introduce significant systematic change
in the column density nor in the intensity of NIR scattered light
produced. Of course, the effects would be stronger if the cloud
included large continuous cavities extending to the centre of the
cloud.

The overall cloud shape was varied by changing the FWHM of the
line-of-sight density distribution by $\pm$30\%. The column densities
were again adjusted according to the 350\,$\mu$m data. When the cloud
was 30\% more extended along the line-of-sight, the resulting column
density towards the model centre was lower by $\sim$20\%. When the
cloud was flattened by 30\%, the percentual increase of the model
column density was closer to 40\%. These numbers do not yet take into
account the fact that a for the best fit to the observed SED also the
intensity of the external radiation field might have to be 
readjusted. This could reduce the net change of the column density.
However, with the radiation field of Fig.~\ref{fig:EA_oh_3.0_1.0} the
160\,$\mu$m surface brightness was already correct to within
$\sim$10\%.

We conclude that the uncertainty of the three-dimensional shape of the
cloud translates to an uncertainty of some tens of per cent in the
column density. These uncertainties affect the accuracy to which the
dust opacity can be determined. However, the modelling is better
constrained when independent extinction measurements are available.

\section{Conclusions}  \label{sect:conclusions}

We have examined the structure of the northern filament of the Corona
Australis cloud using the combination of Herschel sub-millimetre data
and near-infrared observations. The study has lead to the following
conclusions:
\begin{itemize}
\item At $A_{\rm V}<15^{\rm m}$ the three column density estimators,
the thermal dust emission, the NIR reddening of the background stars,
and the surface brightness caused by the NIR scattered light, are all
linearly related to each other. The differences in the actual numbers
can be explained by the uncertainty of the dust opacity and of the
radiation field intensity.
\item At low column densities ($A_{\rm V}\sim 10^{\rm m}$ or below)
the scattered light provides the best column density map in terms of
resolution. Its predictions are very tightly correlated with the
estimates obtained from the dust emission. However, the slope of the
relation depends on the intensity of the radiation field.
\item Based on the dust emission, the north-south column density 
profile was observed to be skewed with a sharp drop in the column
density on the northern side. The profile obtained from the reddening
of the background stars is more symmetric but is strongly affected by
the lack of stars visible through the central filament.
\item The modelling of the dust emission suggests that the radiation
field intensity at the location of the cloud filament is about four
times the value of the normal ISRF. The value can be lowered only by
assuming that the dust sub-millimetre opacity has increased at the
centre of the filament. However, the models of the NIR scattered light
are also consistent with an elevated radiation field intensity.
\item According to the models, the radiation field is anisotropic with
an approximate ratio of 3:1 between the isotropic component and
additional radiation coming from the southern direction.
\item The models are unable to reproduce the deep dip in the NIR
surface brightness at the centre of the filament. One possible
explanation is a change in the dust properties (as indicated by the
differences between the dust models examined) that has lowered the
NIR albedo.
\end{itemize}

\begin{acknowledgements}
MJ acknowledges the support of the Academy of Finland Grants No. 
127015 and 250741.
\end{acknowledgements}

\bibliography{biblio_v2.0}

\begin{thebibliography}{71}
\expandafter\ifx\csname natexlab\endcsname\relax\def\natexlab#1{#1}\fi

\bibitem[{{Anderson} {et~al.}(2010){Anderson}, {Zavagno}, {Rod{\'o}n},
  {Russeil}, {Abergel}, {Ade}, {Andr{\'e}}, {Arab}, {Baluteau}, {Bernard},
  {Blagrave}, {Bontemps}, {Boulanger}, {Cohen}, {Compi{\'e}gne}, {Cox},
  {Dartois}, {Davis}, {Emery}, {Fulton}, {Gry}, {Habart}, {Huang}, {Joblin},
  {Jones}, {Kirk}, {Lagache}, {Lim}, {Madden}, {Makiwa}, {Martin},
  {Miville-Desch{\^e}nes}, {Molinari}, {Moseley}, {Motte}, {Naylor}, {Okumura},
  {Pinheiro Gon{\c c}alves}, {Polehampton}, {Saraceno}, {Sauvage}, {Sidher},
  {Spencer}, {Swinyard}, {Ward-Thompson}, \& {White}}]{Anderson2010}
{Anderson}, L.~D., {Zavagno}, A., {Rod{\'o}n}, J.~A., {et~al.} 2010, \aap, 518,
  L99+

\bibitem[{{Andr{\'e}} {et~al.}(2010){Andr{\'e}}, {Men'shchikov}, {Bontemps},
  {K{\"o}nyves}, {Motte}, {Schneider}, {Didelon}, {Minier}, {Saraceno},
  {Ward-Thompson}, {di Francesco}, {White}, {Molinari}, {Testi}, {Abergel},
  {Griffin}, {Henning}, {Royer}, {Mer{\'{\i}}n}, {Vavrek}, {Attard},
  {Arzoumanian}, {Wilson}, {Ade}, {Aussel}, {Baluteau}, {Benedettini},
  {Bernard}, {Blommaert}, {Cambr{\'e}sy}, {Cox}, {di Giorgio}, {Hargrave},
  {Hennemann}, {Huang}, {Kirk}, {Krause}, {Launhardt}, {Leeks}, {Le Pennec},
  {Li}, {Martin}, {Maury}, {Olofsson}, {Omont}, {Peretto}, {Pezzuto}, {Prusti},
  {Roussel}, {Russeil}, {Sauvage}, {Sibthorpe}, {Sicilia-Aguilar}, {Spinoglio},
  {Waelkens}, {Woodcraft}, \& {Zavagno}}]{Andre2010}
{Andr{\'e}}, P., {Men'shchikov}, A., {Bontemps}, S., {et~al.} 2010, \aap, 518,
  L102

\bibitem[{{Arab} {et~al.}(2012){Arab}, {Abergel}, {Habart}, {Bernard-Salas},
  {Ayasso}, {Dassas}, {Martin}, \& {White}}]{Arab2012}
{Arab}, H., {Abergel}, A., {Habart}, E., {et~al.} 2012, \aap, 541, A19

\bibitem[{{Arzoumanian} {et~al.}(2011){Arzoumanian}, {Andr{\'e}}, {Didelon},
  {K{\"o}nyves}, {Schneider}, {Men'shchikov}, {Sousbie}, {Zavagno}, {Bontemps},
  {di Francesco}, {Griffin}, {Hennemann}, {Hill}, {Kirk}, {Martin}, {Minier},
  {Molinari}, {Motte}, {Peretto}, {Pezzuto}, {Spinoglio}, {Ward-Thompson},
  {White}, \& {Wilson}}]{Arzoumanian2011}
{Arzoumanian}, D., {Andr{\'e}}, P., {Didelon}, P., {et~al.} 2011, \aap, 529,
  L6+

\bibitem[{{Bally} {et~al.}(1987){Bally}, {Lanber}, {Stark}, \&
  {Wilson}}]{Bally1987}
{Bally}, J., {Lanber}, W.~D., {Stark}, A.~A., \& {Wilson}, R.~W. 1987, \apjl,
  312, L45

\bibitem[{{Barnard}(1919)}]{Barnard1919}
{Barnard}, E.~E. 1919, \apj, 49, 1

\bibitem[{{Beckwith} {et~al.}(1990){Beckwith}, {Sargent}, {Chini}, \&
  {Guesten}}]{Beckwith1990}
{Beckwith}, S.~V.~W., {Sargent}, A.~I., {Chini}, R.~S., \& {Guesten}, R. 1990,
  \aj, 99, 924

\bibitem[{{Bergin} \& {Tafalla}(2007)}]{Bergin2007}
{Bergin}, E.~A. \& {Tafalla}, M. 2007, ARA\&A, 45, 339

\bibitem[{{Bernard} {et~al.}(2010){Bernard}, {Paradis}, {Marshall}, {Montier},
  {Lagache}, {Paladini}, {Veneziani}, {Brunt}, {Mottram}, {Martin},
  {Ristorcelli}, {Noriega-Crespo}, {Compi{\`e}gne}, {Flagey}, {Anderson},
  {Popescu}, {Tuffs}, {Reach}, {White}, {Benedetti}, {Calzoletti}, {Digiorgio},
  {Faustini}, {Juvela}, {Joblin}, {Joncas}, {Mivilles-Deschenes}, {Olmi},
  {Traficante}, {Piacentini}, {Zavagno}, \& {Molinari}}]{Bernard2010}
{Bernard}, J.-P., {Paradis}, D., {Marshall}, D.~J., {et~al.} 2010, \aap, 518,
  L88

\bibitem[{{Bohlin} {et~al.}(1978){Bohlin}, {Savage}, \& {Drake}}]{Bohlin1978}
{Bohlin}, R.~C., {Savage}, B.~D., \& {Drake}, J.~F. 1978, \apj, 224, 132

\bibitem[{{Bonnell} {et~al.}(2011){Bonnell}, {Smith}, {Clark}, \&
  {Bate}}]{Bonnell2011}
{Bonnell}, I.~A., {Smith}, R.~J., {Clark}, P.~C., \& {Bate}, M.~R. 2011,
  \mnras, 410, 2339

\bibitem[{{Boulanger} {et~al.}(1996){Boulanger}, {Abergel}, {Bernard},
  {Burton}, {Desert}, {Hartmann}, {Lagache}, \& {Puget}}]{Boulanger1996}
{Boulanger}, F., {Abergel}, A., {Bernard}, J., {et~al.} 1996, \aap, 312, 256

\bibitem[{{Burkert} \& {Hartmann}(2004)}]{Burkert2004}
{Burkert}, A. \& {Hartmann}, L. 2004, \apj, 616, 288

\bibitem[{{Cambr{\'e}sy} {et~al.}(2001){Cambr{\'e}sy}, {Boulanger}, {Lagache},
  \& {Stepnik}}]{Cambresy2001}
{Cambr{\'e}sy}, L., {Boulanger}, F., {Lagache}, G., \& {Stepnik}, B. 2001,
  \aap, 375, 999

\bibitem[{{Carmona} {et~al.}(2007){Carmona}, {van den Ancker}, \&
  {Henning}}]{Carmona2007}
{Carmona}, A., {van den Ancker}, M.~E., \& {Henning}, T. 2007, \aap, 464, 687

\bibitem[{{del Burgo} {et~al.}(2003){del Burgo}, {Laureijs}, {{\'A}brah{\'a}m},
  \& {Kiss}}]{delBurgo2003}
{del Burgo}, C., {Laureijs}, R.~J., {{\'A}brah{\'a}m}, P., \& {Kiss}, C. 2003,
  \mnras, 346, 403

\bibitem[{{D{\'e}sert} {et~al.}(2008){D{\'e}sert}, {Mac{\'{\i}}as-P{\'e}rez},
  {Mayet}, {Giardino}, {Renault}, {Aumont}, {Beno{\^i}t}, {Bernard},
  {Ponthieu}, \& {Tristram}}]{Desert2008}
{D{\'e}sert}, F., {Mac{\'{\i}}as-P{\'e}rez}, J.~F., {Mayet}, F., {et~al.} 2008,
  A\&A, 481, 411

\bibitem[{{Draine}(2003)}]{Draine2003}
{Draine}, B.~T. 2003, \apj, 598, 1017

\bibitem[{{Dupac} {et~al.}(2003){Dupac}, {Bernard}, {Boudet}, {Giard},
  {Lamarre}, {M{\'e}ny}, {Pajot}, {Ristorcelli}, {Serra}, {Stepnik}, \&
  {Torre}}]{Dupac2003}
{Dupac}, X., {Bernard}, J., {Boudet}, N., {et~al.} 2003, A\&A, 404, L11

\bibitem[{{Elmegreen} \& {Elmegreen}(1979)}]{Elmegreen1979}
{Elmegreen}, D.~M. \& {Elmegreen}, B.~G. 1979, \aj, 84, 615

\bibitem[{{Fessenkov}(1952)}]{Fessenkov1952}
{Fessenkov}, V.~G. 1952, Trans. IAU, 8, 707

\bibitem[{{Fischera} \& {Martin}(2012)}]{Fischera2012}
{Fischera}, J. \& {Martin}, P.~G. 2012, ArXiv e-prints

\bibitem[{{Gray} {et~al.}(2003){Gray}, {Corbally}, {Garrison}, {McFadden}, \&
  {Robinson}}]{Gray2003}
{Gray}, R.~O., {Corbally}, C.~J., {Garrison}, R.~F., {McFadden}, M.~T., \&
  {Robinson}, P.~E. 2003, \aj, 126, 2048

\bibitem[{{Griffin} {et~al.}(2010){Griffin}, {Abergel}, {Abreu}, {Ade},
  {Andr{\'e}}, {Augueres}, {Babbedge}, {Bae}, {Baillie}, {Baluteau}, {Barlow},
  {Bendo}, {Benielli}, {Bock}, {Bonhomme}, {Brisbin}, {Brockley-Blatt},
  {Caldwell}, {Cara}, {Castro-Rodriguez}, {Cerulli}, {Chanial}, {Chen},
  {Clark}, {Clements}, {Clerc}, {Coker}, {Communal}, {Conversi}, {Cox},
  {Crumb}, {Cunningham}, {Daly}, {Davis}, {de Antoni}, {Delderfield}, {Devin},
  {di Giorgio}, {Didschuns}, {Dohlen}, {Donati}, {Dowell}, {Dowell}, {Duband},
  {Dumaye}, {Emery}, {Ferlet}, {Ferrand}, {Fontignie}, {Fox}, {Franceschini},
  {Frerking}, {Fulton}, {Garcia}, {Gastaud}, {Gear}, {Glenn}, {Goizel},
  {Griffin}, {Grundy}, {Guest}, {Guillemet}, {Hargrave}, {Harwit}, {Hastings},
  {Hatziminaoglou}, {Herman}, {Hinde}, {Hristov}, {Huang}, {Imhof}, {Isaak},
  {Israelsson}, {Ivison}, {Jennings}, {Kiernan}, {King}, {Lange}, {Latter},
  {Laurent}, {Laurent}, {Leeks}, {Lellouch}, {Levenson}, {Li}, {Li},
  {Lilienthal}, {Lim}, {Liu}, {Lu}, {Madden}, {Mainetti}, {Marliani}, {McKay},
  {Mercier}, {Molinari}, {Morris}, {Moseley}, {Mulder}, {Mur}, {Naylor},
  {Nguyen}, {O'Halloran}, {Oliver}, {Olofsson}, {Olofsson}, {Orfei}, {Page},
  {Pain}, {Panuzzo}, {Papageorgiou}, {Parks}, {Parr-Burman}, {Pearce},
  {Pearson}, {P{\'e}rez-Fournon}, {Pinsard}, {Pisano}, {Podosek}, {Pohlen},
  {Polehampton}, {Pouliquen}, {Rigopoulou}, {Rizzo}, {Roseboom}, {Roussel},
  {Rowan-Robinson}, {Rownd}, {Saraceno}, {Sauvage}, {Savage}, {Savini},
  {Sawyer}, {Scharmberg}, {Schmitt}, {Schneider}, {Schulz}, {Schwartz},
  {Shafer}, {Shupe}, {Sibthorpe}, {Sidher}, {Smith}, {Smith}, {Smith},
  {Spencer}, {Stobie}, {Sudiwala}, {Sukhatme}, {Surace}, {Stevens}, {Swinyard},
  {Trichas}, {Tourette}, {Triou}, {Tseng}, {Tucker}, {Turner}, {Vaccari},
  {Valtchanov}, {Vigroux}, {Virique}, {Voellmer}, {Walker}, {Ward}, {Waskett},
  {Weilert}, {Wesson}, {White}, {Whitehouse}, {Wilson}, {Winter}, {Woodcraft},
  {Wright}, {Xu}, {Zavagno}, {Zemcov}, {Zhang}, \& {Zonca}}]{Griffin2010}
{Griffin}, M.~J., {Abergel}, A., {Abreu}, A., {et~al.} 2010, \aap, 518, L3

\bibitem[{{Hildebrand}(1983)}]{Hildebrand1983}
{Hildebrand}, R.~H. 1983, \qjras, 24, 267

\bibitem[{{Hill} {et~al.}(2011){Hill}, {Motte}, {Didelon}, {Bontemps},
  {Minier}, {Hennemann}, {Schneider}, {Andr{\'e}}, {Men'shchikov}, {Anderson},
  {Arzoumanian}, {Bernard}, {di Francesco}, {Elia}, {Giannini}, {Griffin},
  {K{\"o}nyves}, {Kirk}, {Marston}, {Martin}, {Molinari}, {Nguyen Luong},
  {Peretto}, {Pezzuto}, {Roussel}, {Sauvage}, {Sousbie}, {Testi},
  {Ward-Thompson}, {White}, {Wilson}, \& {Zavagno}}]{Hill2011}
{Hill}, T., {Motte}, F., {Didelon}, P., {et~al.} 2011, \aap, 533, A94

\bibitem[{{Inutsuka} \& {Miyama}(1997)}]{Inutsuka1997}
{Inutsuka}, S.-I. \& {Miyama}, S.~M. 1997, \apj, 480, 681

\bibitem[{{Juvela}(2005)}]{Juvela2005_CRT}
{Juvela}, M. 2005, \aap, 440, 531

\bibitem[{{Juvela} \& {Padoan}(2003)}]{Juvela2003_CRT}
{Juvela}, M. \& {Padoan}, P. 2003, \aap, 397, 201

\bibitem[{{Juvela} {et~al.}(2006){Juvela}, {Pelkonen}, {Padoan}, \&
  {Mattila}}]{Juvela2006_SCA}
{Juvela}, M., {Pelkonen}, V.-M., {Padoan}, P., \& {Mattila}, K. 2006, \aap,
  457, 877

\bibitem[{{Juvela} {et~al.}(2008){Juvela}, {Pelkonen}, {Padoan}, \&
  {Mattila}}]{Juvela2008_CrA}
{Juvela}, M., {Pelkonen}, V.-M., {Padoan}, P., \& {Mattila}, K. 2008, \aap,
  480, 445

\bibitem[{{Juvela} {et~al.}(2009){Juvela}, {Pelkonen}, \&
  {Porceddu}}]{Juvela2009_CrA}
{Juvela}, M., {Pelkonen}, V.-M., \& {Porceddu}, S. 2009, \aap, 505, 663

\bibitem[{{Juvela} {et~al.}(2010){Juvela}, {Ristorcelli}, {Montier},
  {Marshall}, {Pelkonen}, {Malinen}, {Ysard}, {T{\'o}th}, {Harju}, {Bernard},
  {Schneider}, {Vereb{\'e}lyi}, {Anderson}, {Andr{\'e}}, {Giard}, {Krause},
  {Lehtinen}, {Macias-Perez}, {Martin}, {McGehee}, {Meny}, {Motte}, {Pagani},
  {Paladini}, {Reach}, {Valenziano}, {Ward-Thompson}, \&
  {Zavagno}}]{Juvela2010}
{Juvela}, M., {Ristorcelli}, I., {Montier}, L.~A., {et~al.} 2010, \aap, 518,
  L93+

\bibitem[{{Juvela} {et~al.}(2012){Juvela}, {Ristorcelli}, {Pagani}, {Doi},
  {Pelkonen}, {Marshall}, {Bernard}, {Falgarone}, {Malinen}, {Marton},
  {McGehee}, {Montier}, {Motte}, {Paladini}, {Toth}, {Ysard}, {Zahorecz}, \&
  {Zavagno}}]{PaperIII}
{Juvela}, M., {Ristorcelli}, I., {Pagani}, L., {et~al.} 2012, \aap, 541

\bibitem[{{Juvela} \& {Ysard}(2012{\natexlab{a}})}]{Juvela2012_CHI2}
{Juvela}, M. \& {Ysard}, N. 2012{\natexlab{a}}, \aap, 541, A33

\bibitem[{{Juvela} \& {Ysard}(2012{\natexlab{b}})}]{Juvela2012_TB}
{Juvela}, M. \& {Ysard}, N. 2012{\natexlab{b}}, \aap, 539, A71

\bibitem[{{Klessen}(2011)}]{Klessen2011}
{Klessen}, R.~S. 2011, in EAS Publications Series, Vol.~51, EAS Publications
  Series, ed. {C.~Charbonnel \& T.~Montmerle}, 133--167

\bibitem[{{K{\"o}nyves} {et~al.}(2010){K{\"o}nyves}, {Andr{\'e}},
  {Men'shchikov}, {Schneider}, {Arzoumanian}, {Bontemps}, {Attard}, {Motte},
  {Didelon}, {Maury}, {Abergel}, {Ali}, {Baluteau}, {Bernard}, {Cambr{\'e}sy},
  {Cox}, {di Francesco}, {di Giorgio}, {Griffin}, {Hargrave}, {Huang}, {Kirk},
  {Li}, {Martin}, {Minier}, {Molinari}, {Olofsson}, {Pezzuto}, {Russeil},
  {Roussel}, {Saraceno}, {Sauvage}, {Sibthorpe}, {Spinoglio}, {Testi},
  {Ward-Thompson}, {White}, {Wilson}, {Woodcraft}, \& {Zavagno}}]{Konyves2010}
{K{\"o}nyves}, V., {Andr{\'e}}, P., {Men'shchikov}, A., {et~al.} 2010, \aap,
  518, L106

\bibitem[{{Kramer} {et~al.}(2003){Kramer}, {Richer}, {Mookerjea}, {Alves}, \&
  {Lada}}]{Kramer2003}
{Kramer}, C., {Richer}, J., {Mookerjea}, B., {Alves}, J., \& {Lada}, C. 2003,
  \aap, 399, 1073

\bibitem[{{Lehtinen} {et~al.}(2007){Lehtinen}, {Juvela}, {Mattila}, {Lemke}, \&
  {Russeil}}]{Lehtinen2007}
{Lehtinen}, K., {Juvela}, M., {Mattila}, K., {Lemke}, D., \& {Russeil}, D.
  2007, \aap, 466, 969

\bibitem[{{Lombardi} \& {Alves}(2001)}]{Lombardi2001}
{Lombardi}, M. \& {Alves}, J. 2001, \aap, 377, 1023

\bibitem[{{Malinen} {et~al.}(2011){Malinen}, {Juvela}, {Collins}, {Lunttila},
  \& {Padoan}}]{Malinen2011}
{Malinen}, J., {Juvela}, M., {Collins}, D.~C., {Lunttila}, T., \& {Padoan}, P.
  2011, \aap, 530, A101+

\bibitem[{{Marraco} \& {Rydgren}(1981)}]{Marraco1981}
{Marraco}, H.~G. \& {Rydgren}, A.~E. 1981, \aj, 86, 62

\bibitem[{{Mathis} {et~al.}(1983){Mathis}, {Mezger}, \& {Panagia}}]{Mathis1983}
{Mathis}, J.~S., {Mezger}, P.~G., \& {Panagia}, N. 1983, \aap, 128, 212

\bibitem[{{McLeman} {et~al.}(2012){McLeman}, {Wang}, \&
  {Bingham}}]{McLeman2012}
{McLeman}, J.~A., {Wang}, C.~H.-T., \& {Bingham}, R. 2012, ArXiv e-prints

\bibitem[{{Men'shchikov} {et~al.}(2010){Men'shchikov}, {Andr{\'e}}, {Didelon},
  {K{\"o}nyves}, {Schneider}, {Motte}, {Bontemps}, {Arzoumanian}, {Attard},
  {Abergel}, {Baluteau}, {Bernard}, {Cambr{\'e}sy}, {Cox}, {di Francesco}, {di
  Giorgio}, {Griffin}, {Hargrave}, {Huang}, {Kirk}, {Li}, {Martin}, {Minier},
  {Miville-Desch{\^e}nes}, {Molinari}, {Olofsson}, {Pezzuto}, {Roussel},
  {Russeil}, {Saraceno}, {Sauvage}, {Sibthorpe}, {Spinoglio}, {Testi},
  {Ward-Thompson}, {White}, {Wilson}, {Woodcraft}, \&
  {Zavagno}}]{Menshchikov2010}
{Men'shchikov}, A., {Andr{\'e}}, P., {Didelon}, P., {et~al.} 2010, \aap, 518,
  L103

\bibitem[{{Miville-Desch{\^e}nes} {et~al.}(2010){Miville-Desch{\^e}nes},
  {Martin}, {Abergel}, {Bernard}, {Boulanger}, {Lagache}, {Anderson},
  {Andr{\'e}}, {Arab}, {Baluteau}, {Blagrave}, {Bontemps}, {Cohen},
  {Compiegne}, {Cox}, {Dartois}, {Davis}, {Emery}, {Fulton}, {Gry}, {Habart},
  {Huang}, {Joblin}, {Jones}, {Kirk}, {Lim}, {Madden}, {Makiwa}, {Menshchikov},
  {Molinari}, {Moseley}, {Motte}, {Naylor}, {Okumura}, {Pinheiro Gon{\c
  c}alves}, {Polehampton}, {Rod{\'o}n}, {Russeil}, {Saraceno}, {Schneider},
  {Sidher}, {Spencer}, {Swinyard}, {Ward-Thompson}, {White}, \&
  {Zavagno}}]{MAMD2010}
{Miville-Desch{\^e}nes}, M.-A., {Martin}, P.~G., {Abergel}, A., {et~al.} 2010,
  \aap, 518, L104

\bibitem[{{Molinari} {et~al.}(2010){Molinari}, {Swinyard}, {Bally}, {Barlow},
  {Bernard}, {Martin}, {Moore}, {Noriega-Crespo}, {Plume}, {Testi}, {Zavagno},
  {Abergel}, {Ali}, {Anderson}, {Andr{\'e}}, {Baluteau}, {Battersby},
  {Beltr{\'a}n}, {Benedettini}, {Billot}, {Blommaert}, {Bontemps}, {Boulanger},
  {Brand}, {Brunt}, {Burton}, {Calzoletti}, {Carey}, {Caselli}, {Cesaroni},
  {Cernicharo}, {Chakrabarti}, {Chrysostomou}, {Cohen}, {Compiegne}, {de
  Bernardis}, {de Gasperis}, {di Giorgio}, {Elia}, {Faustini}, {Flagey},
  {Fukui}, {Fuller}, {Ganga}, {Garcia-Lario}, {Glenn}, {Goldsmith}, {Griffin},
  {Hoare}, {Huang}, {Ikhenaode}, {Joblin}, {Joncas}, {Juvela}, {Kirk},
  {Lagache}, {Li}, {Lim}, {Lord}, {Marengo}, {Marshall}, {Masi}, {Massi},
  {Matsuura}, {Minier}, {Miville-Desch{\^e}nes}, {Montier}, {Morgan}, {Motte},
  {Mottram}, {M{\"u}ller}, {Natoli}, {Neves}, {Olmi}, {Paladini}, {Paradis},
  {Parsons}, {Peretto}, {Pestalozzi}, {Pezzuto}, {Piacentini}, {Piazzo},
  {Polychroni}, {Pomar{\`e}s}, {Popescu}, {Reach}, {Ristorcelli}, {Robitaille},
  {Robitaille}, {Rod{\'o}n}, {Roy}, {Royer}, {Russeil}, {Saraceno}, {Sauvage},
  {Schilke}, {Schisano}, {Schneider}, {Schuller}, {Schulz}, {Sibthorpe},
  {Smith}, {Smith}, {Spinoglio}, {Stamatellos}, {Strafella}, {Stringfellow},
  {Sturm}, {Taylor}, {Thompson}, {Traficante}, {Tuffs}, {Umana}, {Valenziano},
  {Vavrek}, {Veneziani}, {Viti}, {Waelkens}, {Ward-Thompson}, {White},
  {Wilcock}, {Wyrowski}, {Yorke}, \& {Zhang}}]{Molinari2010}
{Molinari}, S., {Swinyard}, B., {Bally}, J., {et~al.} 2010, \aap, 518, L100

\bibitem[{{Myers}(2009)}]{Myers2009}
{Myers}, P.~C. 2009, \apj, 700, 1609

\bibitem[{{Nguyen Luong} {et~al.}(2011){Nguyen Luong}, {Motte}, {Hennemann},
  {Hill}, {Rygl}, {Schneider}, {Bontemps}, {Men'shchikov}, {Andr{\'e}},
  {Peretto}, {Anderson}, {Arzoumanian}, {Deharveng}, {Didelon}, {di Francesco},
  {Griffin}, {Kirk}, {K{\"o}nyves}, {Martin}, {Maury}, {Minier}, {Molinari},
  {Pestalozzi}, {Pezzuto}, {Reid}, {Roussel}, {Sauvage}, {Schuller}, {Testi},
  {Ward-Thompson}, {White}, \& {Zavagno}}]{NguyenLuong2011}
{Nguyen Luong}, Q., {Motte}, F., {Hennemann}, M., {et~al.} 2011, \aap, 535, A76

\bibitem[{{Ossenkopf} \& {Henning}(1994)}]{Ossenkopf1994}
{Ossenkopf}, V. \& {Henning}, T. 1994, A\&A, 291, 943

\bibitem[{{Ott}(2010)}]{Ott2010}
{Ott}, S. 2010, in Astronomical Society of the Pacific Conference Series, Vol.
  434, Astronomical Data Analysis Software and Systems XIX, ed. Y.~{Mizumoto},
  K.-I. {Morita}, \& M.~{Ohishi}, 139

\bibitem[{{Padoan} \& {Nordlund}(2011)}]{PadoanNordlund2011}
{Padoan}, P. \& {Nordlund}, {\AA}. 2011, \apjl, 741, L22

\bibitem[{{Pagani} {et~al.}(2010){Pagani}, {Steinacker}, {Bacmann}, {Stutz}, \&
  {Henning}}]{Pagani2010}
{Pagani}, L., {Steinacker}, J., {Bacmann}, A., {Stutz}, A., \& {Henning}, T.
  2010, Science, 329, 1622

\bibitem[{{Paradis} {et~al.}(2010){Paradis}, {Veneziani}, {Noriega-Crespo},
  {Paladini}, {Piacentini}, {Bernard}, {de Bernardis}, {Calzoletti},
  {Faustini}, {Martin}, {Masi}, {Montier}, {Natoli}, {Ristorcelli}, {Thompson},
  {Traficante}, \& {Molinari}}]{Paradis2010}
{Paradis}, D., {Veneziani}, M., {Noriega-Crespo}, A., {et~al.} 2010, \aap, 520,
  L8

\bibitem[{{Peterson} {et~al.}(2011){Peterson}, {Caratti o Garatti}, {Bourke},
  {Forbrich}, {Gutermuth}, {J{\o}rgensen}, {Allen}, {Patten}, {Dunham},
  {Harvey}, {Mer{\'{\i}}n}, {Chapman}, {Cieza}, {Huard}, {Knez}, {Prager}, \&
  {Evans}}]{Peterson2011}
{Peterson}, D.~E., {Caratti o Garatti}, A., {Bourke}, T.~L., {et~al.} 2011,
  \apjs, 194, 43

\bibitem[{{Planck Collaboration} {et~al.}(2011{\natexlab{a}}){Planck
  Collaboration}, {Ade}, {Aghanim}, {Arnaud}, {Ashdown}, {Aumont},
  {Baccigalupi}, {Balbi}, {Banday}, {Barreiro}, \& et~al.}]{PlanckII}
{Planck Collaboration}, {Ade}, P.~A.~R., {Aghanim}, N., {et~al.}
  2011{\natexlab{a}}, \aap, 536, A22

\bibitem[{{Planck Collaboration} {et~al.}(2011{\natexlab{b}}){Planck
  Collaboration}, {Ade}, {Aghanim}, {Arnaud}, {Ashdown}, {Aumont},
  {Baccigalupi}, {Balbi}, {Banday}, {Barreiro}, \& et~al.}]{PlanckI}
{Planck Collaboration}, {Ade}, P.~A.~R., {Aghanim}, N., {et~al.}
  2011{\natexlab{b}}, \aap, 536, A23

\bibitem[{{Poglitsch} {et~al.}(2010){Poglitsch}, {Waelkens}, {Geis},
  {Feuchtgruber}, {Vandenbussche}, {Rodriguez}, {Krause}, {Renotte}, {van
  Hoof}, {Saraceno}, {Cepa}, {Kerschbaum}, {Agn{\`e}se}, {Ali}, {Altieri},
  {Andreani}, {Augueres}, {Balog}, {Barl}, {Bauer}, {Belbachir}, {Benedettini},
  {Billot}, {Boulade}, {Bischof}, {Blommaert}, {Callut}, {Cara}, {Cerulli},
  {Cesarsky}, {Contursi}, {Creten}, {De Meester}, {Doublier}, {Doumayrou},
  {Duband}, {Exter}, {Genzel}, {Gillis}, {Gr{\"o}zinger}, {Henning},
  {Herreros}, {Huygen}, {Inguscio}, {Jakob}, {Jamar}, {Jean}, {de Jong},
  {Katterloher}, {Kiss}, {Klaas}, {Lemke}, {Lutz}, {Madden}, {Marquet},
  {Martignac}, {Mazy}, {Merken}, {Montfort}, {Morbidelli}, {M{\"u}ller},
  {Nielbock}, {Okumura}, {Orfei}, {Ottensamer}, {Pezzuto}, {Popesso},
  {Putzeys}, {Regibo}, {Reveret}, {Royer}, {Sauvage}, {Schreiber}, {Stegmaier},
  {Schmitt}, {Schubert}, {Sturm}, {Thiel}, {Tofani}, {Vavrek}, {Wetzstein},
  {Wieprecht}, \& {Wiezorrek}}]{Poglitsch2010}
{Poglitsch}, A., {Waelkens}, C., {Geis}, N., {et~al.} 2010, \aap, 518, L2

\bibitem[{{Schneider} {et~al.}(2010){Schneider}, {Csengeri}, {Bontemps},
  {Motte}, {Simon}, {Hennebelle}, {Federrath}, \& {Klessen}}]{Schneider2010}
{Schneider}, N., {Csengeri}, T., {Bontemps}, S., {et~al.} 2010, \aap, 520, A49

\bibitem[{{Schneider} {et~al.}(2012){Schneider}, {Csengeri}, {Hennemann},
  {Motte}, {Didelon}, {Federrath}, {Bontemps}, {Di Francesco}, {Arzoumanian},
  {Minier}, {Andr{\'e}}, {Hill}, {Zavagno}, {Nguyen-Luong}, {Attard},
  {Bernard}, {Elia}, {Fallscheer}, {Griffin}, {Kirk}, {Klessen}, {K{\"o}nyves},
  {Martin}, {Men'shchikov}, {Palmeirim}, {Peretto}, {Pestalozzi}, {Russeil},
  {Sadavoy}, {Sousbie}, {Testi}, {Tremblin}, {Ward-Thompson}, \&
  {White}}]{Schneider2012}
{Schneider}, N., {Csengeri}, T., {Hennemann}, M., {et~al.} 2012, \aap, 540, L11

\bibitem[{{Schneider} \& {Elmegreen}(1979)}]{Schneider1979}
{Schneider}, S. \& {Elmegreen}, B.~G. 1979, \apjs, 41, 87

\bibitem[{{Shetty} {et~al.}(2009{\natexlab{a}}){Shetty}, {Kauffmann}, {Schnee},
  \& {Goodman}}]{Shetty2009b}
{Shetty}, R., {Kauffmann}, J., {Schnee}, S., \& {Goodman}, A.~A.
  2009{\natexlab{a}}, \apj, 696, 676

\bibitem[{{Shetty} {et~al.}(2009{\natexlab{b}}){Shetty}, {Kauffmann}, {Schnee},
  {Goodman}, \& {Ercolano}}]{Shetty2009a}
{Shetty}, R., {Kauffmann}, J., {Schnee}, S., {Goodman}, A.~A., \& {Ercolano},
  B. 2009{\natexlab{b}}, \apj, 696, 2234

\bibitem[{Skrutskie {et~al.}(2006)Skrutskie, Cutri, Stiening, Weinberg,
  Schneider, Carpenter, Beichman, Capps, Chester, Elias, Huchra, Liebert,
  Lonsdale, Monet, Price, Seitzer, Jarrett, Kirkpatrick, Gizis, Howard, Evans,
  Fowler, Fullmer, Hurt, Light, Kopan, Marsh, McCallon, Tam, Dyk, \&
  Wheelock}]{Skrutskie2006}
Skrutskie, M., Cutri, R., Stiening, R., {et~al.} 2006, \aj, 131, 1163

\bibitem[{{Stamatellos} \& {Whitworth}(2003)}]{Stamatellos2003}
{Stamatellos}, D. \& {Whitworth}, A.~P. 2003, A\&A, 407, 941

\bibitem[{{Steinacker} {et~al.}(2010){Steinacker}, {Pagani}, {Bacmann}, \&
  {Guieu}}]{Steinacker2010}
{Steinacker}, J., {Pagani}, L., {Bacmann}, A., \& {Guieu}, S. 2010, \aap, 511,
  A9+

\bibitem[{{V{\'a}zquez-Semadeni} {et~al.}(2011){V{\'a}zquez-Semadeni},
  {Banerjee}, {G{\'o}mez}, {Hennebelle}, {Duffin}, \&
  {Klessen}}]{VazquezSemadeni2011}
{V{\'a}zquez-Semadeni}, E., {Banerjee}, R., {G{\'o}mez}, G.~C., {et~al.} 2011,
  \mnras, 414, 2511

\bibitem[{{Veneziani} {et~al.}(2010){Veneziani}, {Ade}, {Bock}, {Boscaleri},
  {Crill}, {de Bernardis}, {De Gasperis}, {de Oliveira-Costa}, {De Troia}, {Di
  Stefano}, {Ganga}, {Jones}, {Kisner}, {Lange}, {MacTavish}, {Masi},
  {Mauskopf}, {Montroy}, {Natoli}, {Netterfield}, {Pascale}, {Piacentini},
  {Pietrobon}, {Polenta}, {Ricciardi}, {Romeo}, \& {Ruhl}}]{Veneziani2010}
{Veneziani}, M., {Ade}, P.~A.~R., {Bock}, J.~J., {et~al.} 2010, \apj, 713, 959

\bibitem[{{Wright} {et~al.}(2010){Wright}, {Eisenhardt}, {Mainzer}, {Ressler},
  {Cutri}, {Jarrett}, {Kirkpatrick}, {Padgett}, {McMillan}, {Skrutskie},
  {Stanford}, {Cohen}, {Walker}, {Mather}, {Leisawitz}, {Gautier}, {McLean},
  {Benford}, {Lonsdale}, {Blain}, {Mendez}, {Irace}, {Duval}, {Liu}, {Royer},
  {Heinrichsen}, {Howard}, {Shannon}, {Kendall}, {Walsh}, {Larsen}, {Cardon},
  {Schick}, {Schwalm}, {Abid}, {Fabinsky}, {Naes}, \& {Tsai}}]{Wright2010}
{Wright}, E.~L., {Eisenhardt}, P.~R.~M., {Mainzer}, A.~K., {et~al.} 2010, \aj,
  140, 1868

\bibitem[{{Ysard} {et~al.}(2012){Ysard}, {Juvela}, {Demyk}, {Guillet},
  {Abergel}, {Bernard}, {Malinen}, {M{\'e}ny}, {Montier}, {Paradis},
  {Ristorcelli}, \& {Verstraete}}]{YsardJuvela2012}
{Ysard}, N., {Juvela}, M., {Demyk}, K., {et~al.} 2012, \aap, 542, A21

\end{thebibliography}

\end{document}